\documentclass[prb,superscriptaddress,twocolumn,aps,showpacs,amsmath,amssymb,floatfix]{revtex4-1}

\usepackage{color}
\usepackage{bm}
\usepackage{hyperref}
\usepackage{graphicx}
\usepackage{braket}
\usepackage{MnSymbol}
\usepackage{amssymb}
\hypersetup{colorlinks=true}
\renewcommand{\v}[1]{\textbf{\textit{#1}}}

\begin{document}

\title{Soluble limit and criticality of fermions in $\mathbb Z_2$ gauge theories}

\author{Elio J. K\"onig}
\affiliation{Department of Physics and Astronomy, Center for Materials Theory, Rutgers University, Piscataway, NJ 08854 USA}
\author{Piers Coleman}
\affiliation{Department of Physics and Astronomy, Center for Materials Theory, Rutgers University, Piscataway, NJ 08854 USA}
\affiliation{Department of Physics, Royal Holloway, University of London, Egham, Surrey TW20 0EX, UK}
\author{Alexei M. Tsvelik}
\affiliation{Condensed Matter Physics and Materials Science Division,
Brookhaven National Laboratory, Upton, NY 11973, USA}
\date{\today}

\begin{abstract}

Quantum information theory and strongly correlated electron systems
share a common theme of macroscopic quantum entanglement. In both
topological error correction codes and theories of quantum materials
(spin liquid, heavy fermion and high-$T_c$ systems) entanglement is
implemented by means of an emergent gauge symmetry. Inspired by these
connections, we introduce a simple model for fermions moving in the
deconfined phase of a $\mathbb Z_2$ gauge theory, by coupling Kitaev's
toric code to mobile fermions. This permits us to exactly solve the
ground state of this system and map out its phase diagram.  {Reversing} the sign of the plaquette term in the toric code, {permits us}
to tune the groundstate between an orthogonal metal and an orthogonal
semimetal, in which {gapless quasiparticles survive despite a gap in the spectrum of original fermions.} The small-to-large Fermi surface transition between
these two states occurs in a stepwise fashion with multiple
intermediate phases.  {By using a novel diagrammatic technique we are able to explore physics beyond the integrable point,} to examine various instabilities of the deconfined phase
and to derive the critical theory at the transition between
deconfined and confined {phases}. {We outline how the fermionic toric code can be implemented as a quantum circuit thus providing an important link between quantum materials and quantum information theory. }
\end{abstract}
\date{\today}

\maketitle

\section{Introduction} 
Strongly correlated quantum materials provide natural occurrence of macroscopic entanglement which is believed to be reflected in a variety of exotic experimental observations: the acclaimed Fermi surface reconstruction without symmetry breaking in cuprates~\cite{ProustTaillefer2019} and heavy fermion systems \cite{SiSteglich2010}; quantum oscillations in (bulk) insulating YbB$_{12}$~\cite{XiangLi2018} (and arguably also SmB$_6$~\cite{LiLi2014,TanSebastian2015}); anomalous thermal  transport and spin relaxation in spin liquid candidates, 
e.g.~in the organic salt $\kappa$-ET$_2$Cu$_2$CN$_3$~\cite{YamashitaKanoda2008}. 
All of these materials have the vicinity to (partial) Mott transitions in common (the Kondo breakdown on the lattice can be regarded as an orbital selective Mott localization~\cite{Vojta2010}). 

A theoretically appealing approach to such systems involves fractionalized particles and topological order~\cite{Sachdev2018}. Strong correlations impose (Gutzwiller-) projected local Hilbert spaces. These can be treated in pre-fractionalized slave boson~\cite{Coleman1984} or slave spin~\cite{DeMediciBiermann2005,RueggSigrist2010} theories, whereby a gauge symmetry (typically U(1), SU(2) or $\mathbb Z_2$) is introduced. Topological order enters through the physics of these lattice gauge theories. In particular, sufficiently large space-time dimensions sustain deconfined states, i.e.~macroscopically entangled superposition states with Wegner-Wilson-loops of any length and topological ground state degeneracy on tori. 

Topological order is crucial to explain the Fermi-surface reconstruction without symmetry breaking~\cite{ScheurerSachdev2018}. Conventionally, the Fermi surface volume is fixed by the total electron density~\cite{Luttinger1960,Oshikawa2000} (including f-electrons for Kondo-lattices). However, topological order exploits a loophole~\cite{SenthilVojta2003,ParamekantiVishwanath2004} in the derivation of the Luttinger-Oshikawa theorem. 

\begin{figure}[b]
\includegraphics[scale=.9]{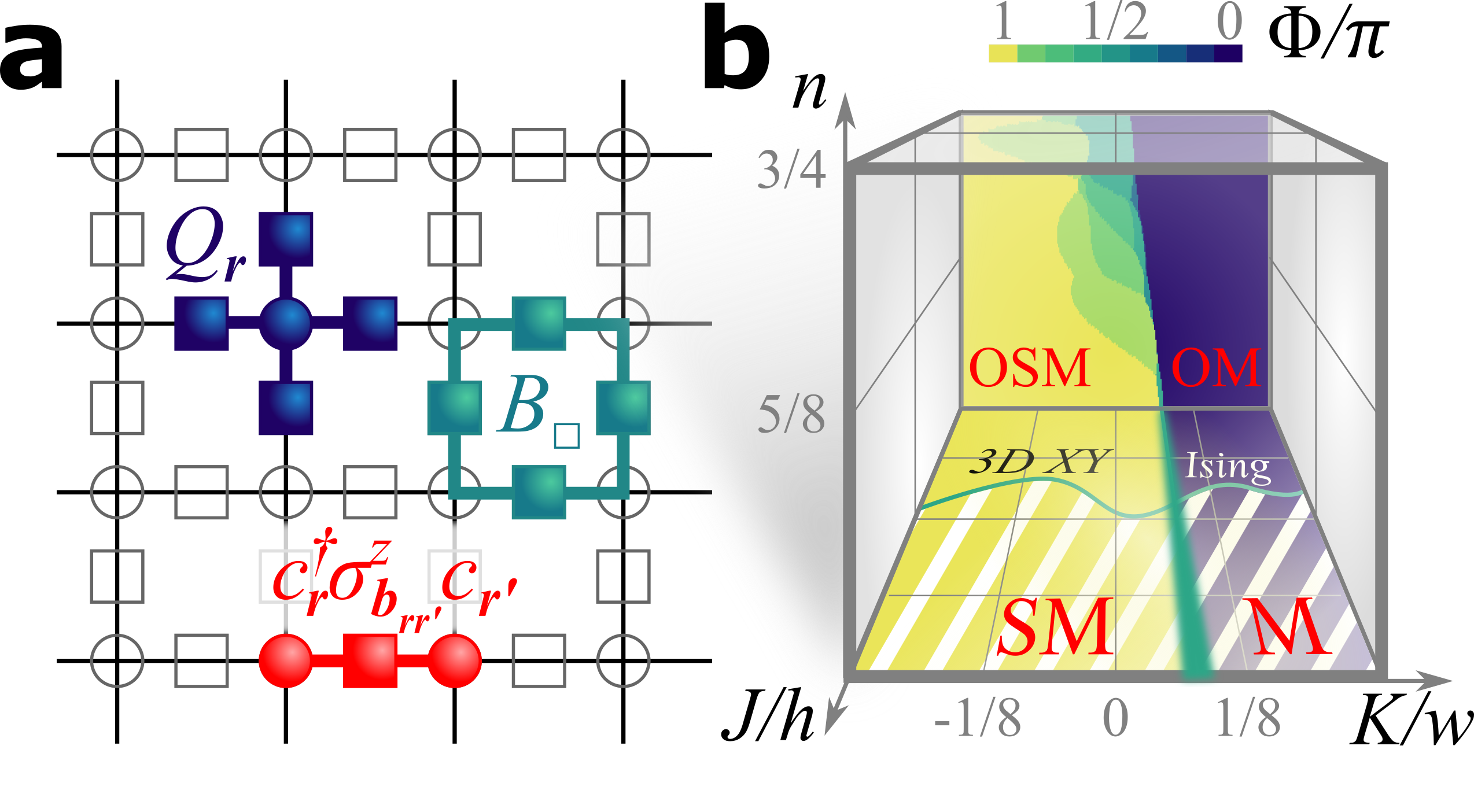}
\caption{{\bf a} Graphical illustration of the mutually commuting
operators in the fermionic toric code, Eq.~\eqref{eq:H0}. {\bf b} Phase diagram [(O)M = (orthogonal) metal, (O)SM = (orthogonal) semimetal] as a function of coupling constants $K/w$,  {$J/h$} and filling $n$. {Numerical data in the $J = 0$ plane (in the back) is combined with a schematic illustration for $J >0$.}  }
\label{fig:model}
\end{figure}

The same 
macroscopic entanglement associated with topological order is also utilized in quantum error correction codes. For example, Kitaev's soluble toric code model~\cite{Kitaev1997} interweaves numerous imperfect physical qubits to two robust logical qubits. We here exploit this insight from quantum information science and expose the toric code to a fermionic bath,~Fig.~\ref{fig:model} {\bf a}: We thereby obtain asymptotically exact analytical results about deconfined states of gauge theories coupled to itinerant electrons. {Additionally, we extend previous toric-code-proposals to design an analogue quantum computer of fermionic $\mathbb Z_2$ gauge theories, Fig.~\ref{fig:ToricCodeExperiment}.} 

Recently, there has been substantial numerical progress in the study of deconfinement in metals~\cite{AssaadGrover2016,GazitVishwanath2017,GazitWang2018,HofmannGrover2019,ChenMeng2019,GazitSachdev2019,ChenMeng2020}. Certain fermionic $\mathbb Z_2$ gauge theories are amenable to Quantum-Monte-Carlo methods (sign-free) and provide evidence for small-to-large Fermi surface transitions without symmetry breaking~\cite{ChenMeng2019,GazitSachdev2019,ChenMeng2020}. Despite this, numerically realistic system sizes and the pertinent obstacle of analytical continuation in frequency space are still a limitation in resolving sharp Fermi surface features. Complementary techniques which overcome such problems, in particular simple analytically tractable models of fermions in deconfined gauge theories~\cite{KoenigKomijani2020} are widely lacking. A promising approach~\cite{SeifertVojta2018,ChoiKim2018} is to Kondo couple conduction fermions to the simplest exactly soluble spin liquid with deconfined $\mathbb Z_2$ gauge degrees of freedom - Kitaev's honeycomb model~\cite{Kitaev2006}. However, to the best of our knowledge, only perturbative or mean field results are available to date. 
  
In this {article} we introduce a simple model of fermions in the deconfined (i.e. topological) phase of $\mathbb Z_2$ gauge theory - a fermionic toric code~\cite{Kitaev1997,LevinStern2011,ZhongLuo2013,GuWen2014,ProskoMaciejko2017}, see Eq.~\eqref{eq:H0} below and Fig.~\ref{fig:model}~{\bf a}. 
We {begin}
 our discussion from the asymptotic cases $K \gg w$ [$K \ll - w$] in the phase diagram~\ref{fig:model}~{\bf b}, when the ground state is easily determined to be an orthogonal metal (OM) [orthogonal semimetal (OSM)] with large [small] Fermi surface. We are able to characterize the small to large Fermi surface transition as an infinite sequence of symmetry broken states with fractional average flux $\Phi$ and develop a diagrammatic technique to systematically include perturbations about this soluble point and to study the transition to the confined phase.

The {OM concept} was introduced in Ref.~\onlinecite{NandkishoreSenthil2012} as a state similar to a normal metal in all respects (e.g. conductivity and thermodynamics) except for the behavior of single electron Green's functions (e.g. the spectral function is gapped).  In~\onlinecite{NandkishoreSenthil2012} the lattice fermions were fractionalized into ``orthogonal'' fermions and slave spins
$
c_{{\bf r},\alpha} \rightarrow \tau^z_{{\bf r}} f_{{\bf r},\alpha},$
where $\tau^a_{\v r}$ are Pauli matrix operators. In the OM, the $\tau$-spins are disordered, while the $f$-fermions are in a Fermi liquid state. The authors of~\onlinecite{NandkishoreSenthil2012} provided exemplary solvable models. 
 Recently the authors of~\onlinecite{GazitSachdev2019} have introduced a model of OM as a $\mathbb Z_2$ gauge theory where $\tau$-spins played a role of Higgs bosons. 

Here we suggest a radical simplification of the theory by generalizing the mapping between the toric code and $\mathbb Z_2$ gauge theory~\cite{Wegner1971,Tupitsyn2010}. {The explicit application of Gauss' law} removes the necessity of $f$-fermions and Higgs bosons and enables us to work with gauge invariant fermions (see~App.~\ref{app:IsingHiggs} for details) which extends our calculational capacities. The exact solution of the resulting fermionic toric code and associated diagrammatics play the same role as free fermions in ordinary metals providing the starting point for perturbation theory and a {positive} definition of a ``($\mathbb Z_2$-deconfined) non-Fermi liquid'' as the class of quantum states of non-integrable models which are adiabatically connected to the ground state of the soluble model.

\section{Bare model}
The soluble starting point for our discussions is a generalization of Kitaev's toric code~\cite{Kitaev1997} by means of fermionic matter fields. The basic Hamiltonian $H_0 = H_K + H_h + H_c $ is given by
\begin{subequations}
\begin{eqnarray}
H_K &=& - K \sum_{\square} B_\square, \quad H_h = - h \sum_{\v r} Q_{\v r}, \\
H_c &=& - w \sum_{\langle \v r, \v r' \rangle}  \sigma^z_{\v b_{\v r, \v r'}}c^\dagger_{\v r, \alpha} c_{\v r', \alpha}  - \mu \sum_{\v r} c^\dagger_{\v r, \alpha} c_{\v r, \alpha}.
\end{eqnarray}
\label{eq:H0}
\end{subequations}
Here, $c^\dagger_{\v r,\alpha}$ creates a fermion with spin component $\alpha = \uparrow, \downarrow$ at a vertex $\v r$ of a square lattice ({depicted} by circles in Fig.~\ref{fig:model} {\bf a}), while $\mathbb Z_2$ gauge fields are represented by Pauli matrices $\sigma^a_{\v b}$ ($a = x,y,z$) located on each bond $\v b$  ({depicted} by squares in Fig.~\ref{fig:model} {\bf a}). The flux (= plaquette) operators $B_\square = \prod_{\v b \in \square} \sigma_{\v b}^z$ and charge (= star) operators $Q_{\v r} = (-1)^{\hat n_r} \prod_{\v b \in +_{\v r}} \sigma_{\v b}^x$ (where $\hat n_{\v r} = \sum_\alpha c^\dagger_{\v r,\alpha}c_{\v r,\alpha}$) all mutually commute, and moreover commute with the fermionic term $H_c$ (we assume $w>0$). 
In distinction to a model studied in Refs.~\onlinecite{ProskoMaciejko2017,ZoharCirac2017,SmithKovrizhin2018,FuPerkins2018},   in Eq.~\eqref{eq:H0} a factor $(-1)^{\hat n_{\v r}}$  is included into $Q_{\v r}$ which allows the following projective construction of the ground state.  

\subsection{Ground states for $\vert K/w \vert \gg 1$} As in the toric code, the construction of the ground state $\ket{\rm GS}$ of Eq.~\eqref{eq:H0} relies on an extensive number of integrals of motion $B_\square$, $Q_{\v r}$ with eigenvalues $\pm 1$. 
We first consider {the limit $\vert K/w\vert \gg 1$ in which all ground states are homogeneous} with zero flux ($\pi$-flux), $B_{\square} \ket{\rm GS_0} = \ket{\rm GS_0}$ ($B_{\square} \ket{\rm GS_\pi} = -\ket{\rm GS_\pi}$), through all plaquettes. {It is illustrative to first} set $h = 0$. In this limit $\sigma^z_{\v b}$ are classical variables and we choose a gauge in which the {gauge} sector of the 0-flux ($\pi$-flux) solution, denoted $\ket{0}_\sigma$ ($\ket{\pi}_\sigma$), suffices $\langle  \sigma^z_{\v b} \rangle = 1$ ($\langle  \sigma^z_{\v b} \rangle = (-1)^{b_x}$). Then, the fermionic term $H_c$ can be readily solved by Fourier transform. Of course, the dispersion is different in the 0-flux [$\epsilon_0(\v k) = -w (\cos (k_x) + \cos(k_y)); \,\v k \in (- \pi, \pi)\times (- \pi, \pi)$] and $\pi$-flux background [$\epsilon_\pi^\pm(\v k) = \pm w \sqrt{\cos^2 (k_x) + \cos^2(k_y)}; \, \v k \in (-\pi/2, \pi/2) \times (\pi,\pi)$]. In either case, the ground state in the fermionic sector is a Fermi sea which we denote $\ket{{\rm FS}_{0/\pi}}_c$. The re-imposition of $h > 0$ {requires} $Q_{\v r} \ket{\rm GS} = \ket{\rm GS} (\forall \v r)$. This lifts the macroscopic degeneracy of {candidate} ground states leaving only two contenders in the infinite plane
\begin{subequations}
\begin{align}
\ket{\rm{GS}_{0}} &= \prod_{\v r} \hat P_{\v r} \left [ \ket{{\rm FS}_0}_c  \ket{0}_\sigma \right ],\; E_0 = - K -  h  + 2 E_{c,0}, \\
 \ket{\rm{GS}_{\pi}} &= \prod_{\v r} \hat P_{\v r} \left [ \ket{{\rm FS}_\pi}_c  \ket{\pi}_\sigma \right ],\; E_\pi = K -  h  + 2 E_{c,\pi},
\end{align}
\label{eq:Solutions}
\end{subequations}
where $\hat P_{\v r} = (1 + Q_{\v r})/2 = \hat P_{\v r}^2$.
These two states represent superpositions of configurations of $\sigma$-fields which preserve the flux configuration. The fermion dispersion{, denoted $\epsilon_0(\v k)$ [$\epsilon_\pi(\v k)$],} of these phases enable us to identify $\ket{GS_0}$ [$\ket{GS_\pi}$] as an OM [OSM], respectively.
We emphasize that, despite the inhomogenous {gauge field} configuration and the small semimetallic Fermi surface,  $\ket{GS_\pi}$ breaks neither  crystalline symmetries (the latter being projectively represented, see App.~\ref{app:Symmetries}) 
nor the Oshikawa-Luttinger theorem (because the gauge sector is deconfined, see App.~\ref{app:LuttingerTheorem})~\cite{ParamekantiVishwanath2004}.

\subsection{Small to large Fermi surface transition}
\label{sec:OSMOMTransition}

While it is clear that the eigenstates presented in Eq.~\eqref{eq:Solutions} yield the correct ground state for $\vert K /w \vert \gg 1$, inhomogenous states displaying arrays of $\pi$ fluxes with density $\Phi \neq 0, \pi$ become important at small $\vert K/w \vert$. These are not favorable for $H_K$, but yield energetic gain of order $w$ by lowering the ground state of the electrons. The latter effect is especially great when the band is at commensurate (e.g.~half-)filling due to the nesting of the Fermi surface. Indeed~\cite{ProskoMaciejko2017}, Monte Carlo simulations corroborate the conjecture that the average flux density at $K = 0$ and filling $n$ is $\Phi = 2\pi n$. {The situation is somewhat similar to the quantum Hall effect with its succession of various quantum Hall states.}~\cite{HasegawaWiegmann1989}

{For a given flux density $\Phi$, it is reasonable to assume that the
ground state configuration is given by some regular array
of $\pi$-fluxes. For each of these,} a ground state $\ket{\text{GS}_\Phi} = \prod_{\v r} \hat P_{\v r} [\ket{\text{FS}_\Phi}_c \ket{\Phi}_\sigma]$ can be readily constructed and the ground state energy determined. Contrary to the OM and OSM, the {intermediate} states do break crystalline symmetries, even when represented projectively. This is revealed in the ground state averages of the $B_{\square}$-operators which are invariant under action of the projection operators. Therefore, we expect that the OM-OSM transition, which separates distinct gapless quantum phases characterized by different projective representation of translations~\cite{GazitSachdev2019}, occurs as an infinite succession of symmetry broken states with fractional average flux $\Phi = \pi k/N_\Phi$ ($k = 1, \dots, N_\Phi-1$). 

\begin{figure}
\includegraphics[width = .45\textwidth]{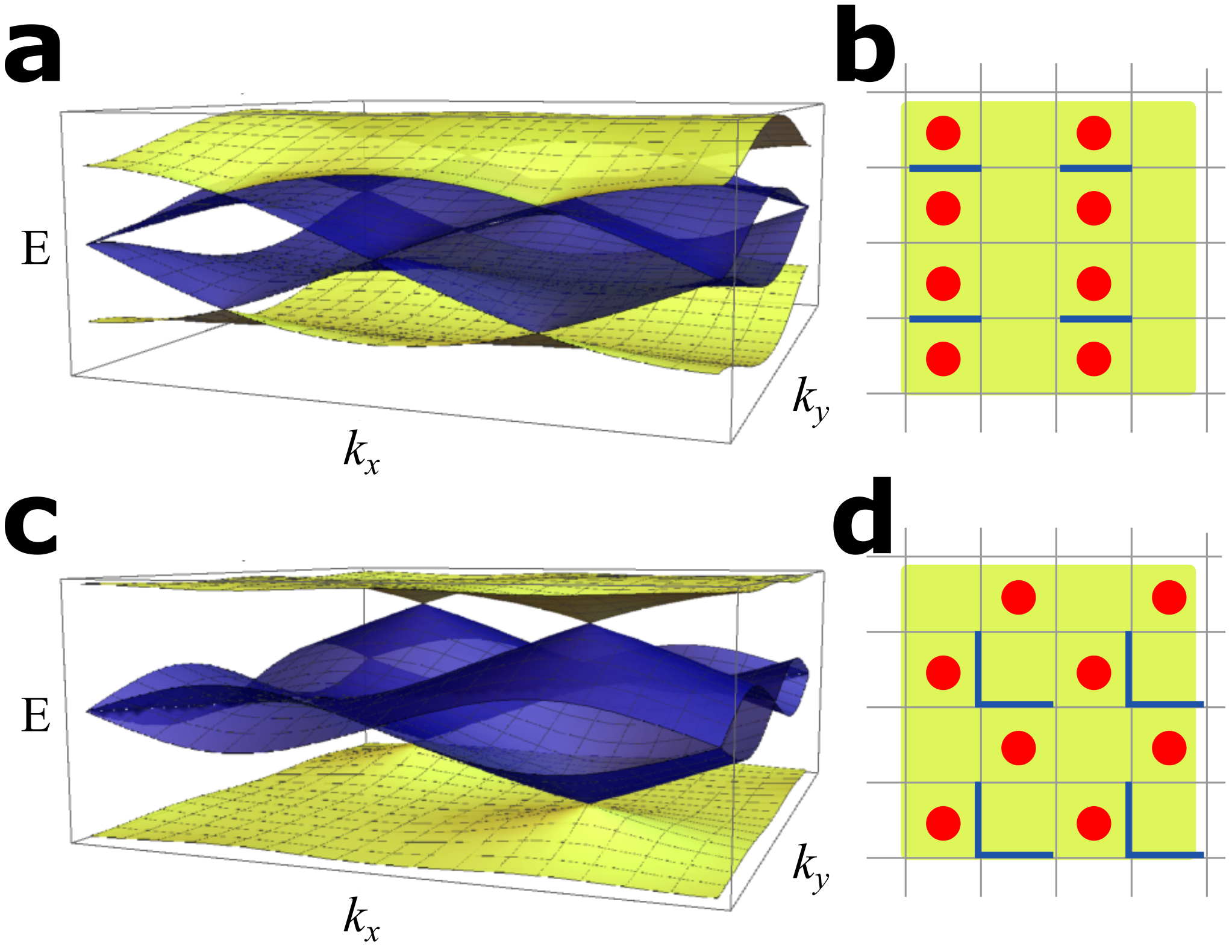}
\caption{{Band structure associated to flux configurations with average flux $\Phi = \pi/2$ and (a) stripy and (c) checkerboard arrangement of fluxes. Within a unit cell (yellow square), the latter are presented in (b,d): a $\pi$ flux is depicted by a red dot, we choose a gauge in which $\text{sign}(w)$ is reversed on highlighted bonds.}}
\label{fig:Pi2Disp}
\end{figure}

To substantiate this hypothesis, we have semi-analytically investigated a large variety of trial flux configurations for $N_\Phi = 8$. {While we relegate technical details to~App.~\ref{app:OSMOMTransition}, 
%
we here illustrate the procedure and consider two exemplary states with average flux $\Phi = \pi/2$: an arrangement of vertical lines and a checkerboard pattern. The corresponding eigenvalues $E$ are determined by  
\begin{subequations}
\begin{eqnarray}
&4&-\cos (4 k_x)+8 \left(E ^2-2\right) E ^2+\left(4-8 E ^2\right) \cos (2 k_y)\notag\\ &&+\cos (4 k_y)= 0 \quad\text{ (vertical stripes),} \\
&2&-4 \sin (2 k_x) \sin (2 k_y)-\cos (4 k_x)+8 \left(E ^2-2\right) E ^2\notag\\ &&-\cos (4 k_y)= 0 \quad\text{ (checkerboard).} 
\end{eqnarray}
\end{subequations}
The configurations and corresponding band structures are presented in Fig.~\ref{fig:Pi2Disp}. In particular, for the checkerboard pattern there are 4 Dirac points in the Brillouin zone. The ground state energy as a number of filling is readily obtained for these two configurations and within the $\pi/2$ sector, the stripy (checkerboard) pattern is favorable near half (quarter) filling, see App.~\ref{app:OSMOMTransition}. We repeat the procedure for $\sim 30$ other trial states and represent} the flux density { associated} with lowest energy as a color plot in Fig.~\ref{fig:model}~{\bf b}. As the stepsize $1/N_\Phi \rightarrow 0$, {the observed succession of states} is expected to coalesce into a quantum phase transition with a finite, critical, strange metallic region. 

\begin{figure}[t]
\includegraphics[scale=.55]{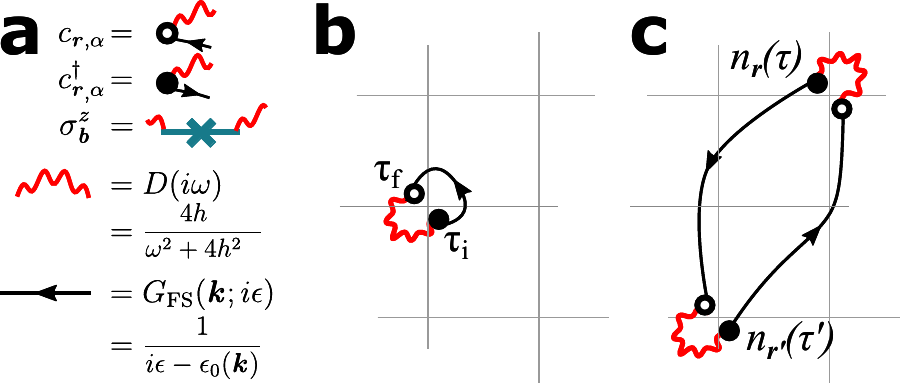}
\caption{{\bf a} Diagrammatic {representation of operators} and bare propagators (here for the OM phase). {\bf b} While the fermionic Green's function is gapped, operator insertions with an even number of fermionic fields at the same space-time position (e.g. density) display ordinary Fermi-liquid behavior, because $D(\tau = 0)= 1$. {\bf c} This holds in particular for the polarization operator.}
\label{fig:Correlators}
\end{figure}

\subsection{Excitations} We return to the OM and OSM phases, for which fermionic single particle excitations are 
\begin{subequations}
\begin{align}
\ket{e:\v k} &= \prod_{\v r} \hat P_{\v r} [c_{\v k}^\dagger \ket{{\rm FS}_{0/\pi}}_c\ket{{0/\pi}}_\sigma] , & \v k \not{\in} \text{ Fermi sea,} \\
\ket{h:\v k} &= \prod_{\v r} \hat P_{\v r} [c_{\v k} \ket{{\rm FS}_{0/\pi}}_c\ket{{0/\pi}}_\sigma], &  \v k \in \text{ Fermi sea.}
\end{align}
\label{eq:FermionicExcitations}
\end{subequations}
Electrons (holes) have excitations energy $\epsilon_{0/\pi}(\v k) - \mu$ ($\mu -\epsilon_{0/\pi}(\v k)$) above the ground state. Particle-hole pairs and multifermion excitations can be expressed analogously by insertion of fermionic operators to the right of all projectors $\hat P_{\v r}$. {In contrast to these gapless excitations, states obtained by applying fermionic operators to the left of projectors are gapped, because fermion operators anticommute with $Q_{\v r}$ and thus create a local excitation with energy $2h$.} Electric strings $W_{\gamma_{\v r, \v r'}}^{(e)} = \prod_{\v b \in \gamma_{\v r, \v r'}} \sigma_{\v b}^z$ along a contour $\gamma_{\v r, \v r'}$ also create {the same local} excitations. As in the toric code, {strings} are deconfined and have energy 2h at each end {(this motivates the notion of ``e''-particles)}. However, {unlike} the toric code, magnetic strings $W_{\gamma^*_{\square, \square'}}^{(m)} = \prod_{\v b \in \gamma^*_{\square, \square'}} \sigma_{\v b}^x$ along a dual contour $\gamma^*_{\square, \square'}$ do not create static eigenstates. 

\begin{figure}
\includegraphics[scale=.55]{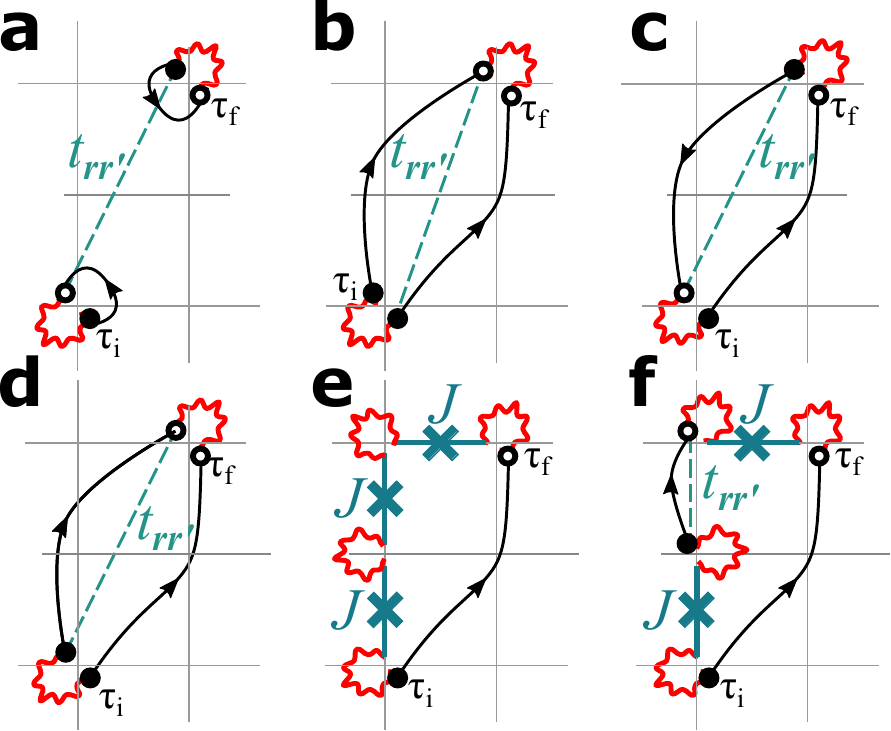}
\caption{Diagrammatic representation of perturbative  $t$- and $J$-corrections to $G(\v r_f, \v r_i; \tau_f,\tau_i)$. {\bf a}-{\bf d} First order diagrams in $t$. {\bf e}-{\bf f} Diagrams contributing to the {Higgs} transition.}
\label{fig:Perturbations}
\end{figure}

\subsection{Diagrammatic technique} Despite the absence of a Wick theorem {in the {toric code} sector,} the presence of Wick's theorem for fermions in a Fermi sea allows us to develop a Feynman diagrammatic representation of imaginary time {ordered} ground state correlators of fermionic operators and of $\sigma^z$ insertions (see Fig.~\ref{fig:Correlators} and~App.~\ref{app:Diagram}). {In distinction to ordinary diagrammatics, a fermion operator inserts a vertex of a local gapped propagator and a dispersive, orthogonal fermion. This is because $c^\dagger_{\v r}$ and $c_{\v r}$ create an ``e'' particle in addition to a fermionic excitation. Similarly, $\sigma^z_{\v b}$ inserts a vertex connecting two ``e'' particles at sites adjacent to the bond $\v b$. The absent Wick theorem implies non-trivial (but computable) interactions of ``e'' particles on sites with more than two vertices.}

The simplest correlator - the two point Green's function, Fig.~\ref{fig:Correlators}~{\bf b} - is $G(\v r_1, \v r_2; \tau ) = \delta_{\v r_1, \v r_2} e^{- 2h \vert \tau\vert} G_{\rm FS}(\v r_1, \v r_1; \tau)$ or in frequency domain
 \begin{equation}
G(\v r_1, \v r_1; z) =  \int (dk) \frac{1}{z - \epsilon_0(\v k) + \text{sign}[\epsilon_0(\v k)]2h}.
\end{equation}
This determines {a} gapped density of states.
On the other hand, the correlators of local two-fermion operators
(e.g. the polarization operator) display standard Fermi-liquid
behavior, Fig.~\ref{fig:Correlators}~{\bf c}. Thus 
Eq.~\eqref{eq:H0} provides a realization of an orthogonal
metal~\cite{NandkishoreSenthil2012}. As a corollary, the instability
of the OM (OSM) with respect to any fermionic interaction with local
space-time operators, $H_{\rm int} = \sum_{\v r, \v r'} c^\dagger_{\v
r, \alpha} c_{\v r, \beta} c^\dagger_{\v r', \alpha'} c_{\v r',
\beta'} V_{\alpha, \beta; \alpha', \beta'}(\v r, \v r')$, is exactly
the same as in the corresponding confining (i.e.~trivial) Fermi liquid
phase.

While we have followed the current convention~\cite{NandkishoreSenthil2012,GazitVishwanath2017,GazitSachdev2019}, by considering $c_{\v r}^\dagger$ as the creation operator of the physical fermion, following Dirac~\cite{Dirac1927} we could equally have identified $\tilde c^\dagger_{\v r} = W_{\gamma_{\infty, \v r}}^{(e)} c^\dagger_{\v r}$ as the physical creation operator. This operator simultaneously creates fermions and the associated distortion in the gauge field. It is this object that creates the gapless excitations in Eq.~\eqref{eq:FermionicExcitations}.

\section{Perturbation theory} The diagrammatic technique allows the systematic study of perturbations which break local charge conservation $[\delta H, Q_{\v r}] \neq 0$,
\begin{equation}
\delta H = - \sum_{\v r, \v r'} t_{\v r, \v r'} c^\dagger_{\v r, \alpha} c_{\v r', \alpha} - J\sum_{\v b} \sigma^z_{\v b}. \label{eq:Hpert}
\end{equation}
{Perturbative} contributions in $t_{\v r, \v r'}$ (represented by a dashed line) to the Green's function are depicted in Fig.~\ref{fig:Perturbations}{, which also illustrates the string tension $J$ of electric strings (we leave a finite string tension of magnetic strings for future studies).} {In the dual formulation of an Ising-Higgs gauge theory, $J$ represents the nearest neighbor $\mathbb Z_2$ slave spin interaction.}

\subsection{Long-range hopping}

An infinite order resummation of the hopping in the random phase approximation (RPA) becomes justified for long-range $t_{\v r, \v r'}$: In all diagrams except of Fig.~\ref{fig:Perturbations}~{\bf a}, $t_{\v r_i,\v r_f}$ is multiplied by the Green's functions connecting the same sites. The decay of $G_{\rm FS}(\v r_f, \v r_i; \tau_f, \tau_i)$ in space removes the singularity of the Fourier transform in momentum space. This validates the omission of this kind of diagrams in RPA and thus,
\begin{equation}
G_{\rm RPA}(\v k, z )  = [G(\v x, \v x; z )^{-1} + t(\v k)]^{-1}.
\end{equation}
This implies the appearance of dispersive subgap states at energy $E(\v k)$. For example, for the OM phase and constant density of states, $G(\v x, \v x; z) = \rho_{0} \ln[(2h - z)/(2h + z)]$ and thus $E(\v k) = 2h \tanh[1/(\rho_0 t(\v k)]$.

\begin{figure}[t]
\includegraphics[scale=.55]{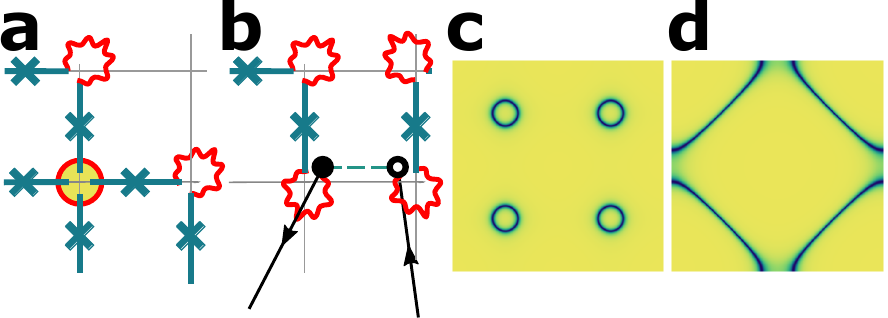}
\caption{{\bf a} ({\bf b}) Interaction of order parameter field with itself (with fermions) {(4-point correlators of ``e''-particles are represented as a disk)}. {\bf c} ({\bf d}) Spectral weight in the extended Brillouin zone in the confined semimetallic (confined metallic) phase at 10\% doping above half-filling {and elastic scattering rate $1/\tau = 0.2 w$}.}
\label{fig:NonLinFS}
\end{figure}

\subsection{{Higgs} transition ($K \gg w$)} According to {the Feynman rules}, fermionic operators are glued together with ``e'' particles (i.e. the ends of $W^{(e)}_{\gamma_{\v r,\v r'}}$ strings) in the deconfining phase of the toric code. Technically, this is reflected in the Green's function obtained by the resummation of diagrams of type Fig.~\ref{fig:Perturbations}~{\bf e}
\begin{equation}
 G(\v r_f, \v r_i; \tau_f, \tau_i) = \mathbf{D}(\v r_f, \v r_i; \tau_f, \tau_i) G_{\rm FS}(\v r_f, \v r_i; \tau_f, \tau_i).\label{eq:GTOT}
 \end{equation}
When $J \neq 0$, strings $\mathbf{D}(\v r_f, \v r_i; \tau_f, \tau_i)$ are non-zero even for $\v r_f \neq \v r_i$. They describe the dynamics of ``e'' particles and lead to finite $\langle \sigma^z_{\v b}(\tau) \sigma^z_{\v b'} (\tau')\rangle$ correlators. We first concentrate on the OM phase, where the propagator of ``e'' particles is determined self-consistently to be
\begin{equation}
\mathbf D(\v q; i \omega) = \frac{4h}{\omega^2 + 4h (h - 2 J [\cos(q_x) + \cos(q_y)])} \label{eq:DResum}.
\end{equation}
The inclusion of small nearest neighbor hopping $t$ leads to the replacement $J \rightarrow J + \bar t $ in Eq.~\eqref{eq:DResum}, see Fig.~\ref{fig:Perturbations}~{\bf f}, where $\bar t = 2 t  G_{\rm FS}(\v r+\hat e_x, \v r;\tau, \tau)$. At small $J,\bar t$ the intersite Green's function is finite, but exponentially suppressed.

The zero frequency, zero momentum correlator $\mathbf D(\v q= 0; i\omega =0)$ represents the sum over electric strings of any spatiotemporal extent. Its divergence at $4(J + \bar t)=h$ signals the confinement-deconfinement quantum phase transition of the toric code~\cite{FradkinShenker1979,TrebstNayak2007,VidalSchmidt2009,Tupitsyn2010}. For even larger $J+\bar t$, the {condensation} of {``e''-particles} imposes the breakdown of topological order and the propagator is $\mathbf D(\v q; i\omega) = Z (2\pi)^3 \delta(\v q) \delta(\omega) + \delta \mathbf D(\v q; i\omega)$. According to Eq.~\eqref{eq:GTOT}, fermions form an ordinary Fermi liquid, for which the toric code order parameter $\sqrt{Z}$ determines the fermionic quasiparticle weight.

To determine the behavior near criticality, one has to incorporate {renormalization} corrections to the strings and fermionic propagators in Fig.~\ref{fig:Perturbations}~{\bf e},{\bf f}. This is most systematically achieved within an effective field theory $S = S_\phi + S_\psi + S_{\rm int}$ with
\begin{subequations}\label{eq:EffFieldTheoryOM}
\begin{align}
S_\phi &= \int d\tau d^2x \;\frac{1}{2} \phi [- \partial_\tau^2 - v^2 \nabla^2 + r] \phi + \frac{\lambda}{4 } \phi^4 ,\label{eq:Phi4} \\
S_\psi &= \int d\tau d^2x \;\bar\psi_\alpha [\partial_\tau + \epsilon_0(-i \partial_x,-i \partial_y)] \psi_\alpha ,\\
S_{\rm int} &=  \int d\tau d^2x \; g \phi^2  \bar \psi_\alpha [\cos(-i \partial_x)+\cos(-i \partial_y)]  \psi_\alpha \label{eq:Sint}.
\end{align}
\end{subequations}
The neutral field $\phi$ ($\psi$) describes the critical fluctuation of strings (of the fermions), $\mathbf D(\v x; \tau) = 4 h a^2 \langle \phi(\v x, \tau) \phi(0,0) \rangle$ ($G_{\rm FS}(\v x; \tau) = a^2 \langle \bar \psi(\v x, \tau) \psi(0,0) \rangle$), where $a$ is the lattice constant. We can thus identify  $r = 4h (h - 4 (J + \bar t))$, $v^2 = 4h (J + \bar t) a^2$. Moreover, we determined the coupling constants $\lambda \sim a^2 J^4/h$ (cf.~Fig.~\ref{fig:NonLinFS}~{\bf a}) and $g \sim a^2 ht $ (cf.~Fig.~\ref{fig:NonLinFS}~{\bf b} and~App.~\ref{app:ConfDeconfTransition}, note the formfactor in Eq.~\eqref{eq:Sint} due to nearest neighbor fermionic insertions) and we reiterate that this field theory is designed to incorporate perturbations on top of the integrable theory, hence $g \propto t$. As an important corollary of the microscopic derivation, {at $t = 0$ the presence of fermions does not affect the} 3D Ising {criticality, yet it is a relevant perturbation}~\cite{SachdevMorinari2002,NandkishoreSenthil2012,ZhongLuo2012}. 
The critical theory Eq.~\eqref{eq:EffFieldTheoryOM}, can then be used to determine a variety of critical properties at the {Higgs} 
transition. Most prominently, the quasiparticle weight plays the role of the order parameter, i.e. $Z \sim \vert h^2 r /J^4 \vert^{2\beta}$ where 
$\beta \approx 0.33$ [(2+1)D Ising]. 

\subsection{{Higgs} transition ($K \ll -w$)}

We now return to the {Higgs}  
transition induced by
perturbing the OSM phase with Eq.~\eqref{eq:Hpert}. Conceptually, the same
steps which we outlined for the OM hold in the OSM case, too. However,
the non-trivial representation of translational symmetry in the
flux-phase implies several subtleties [see~App.~\ref{app:ConfDeconfTransition} for details
including a derivation of the OSM analogue of
Eq.~\eqref{eq:EffFieldTheoryOM}]: (i) The propagator $\mathbf D(\v r_f, \v
r_i; \tau_f, \tau_i)$ is a matrix which acts in the space of the basis
of the two-atomic unit cell. (ii) As a consequence, the transition
occurs at a slightly higher numerical value of $(J+\bar t)/h =
1/\sqrt{8}$ and there are two momenta $\v q$ in the Brillouin zone, at
which $\mathbf D(\v q, i \omega = 0)$ diverges. (iii) The relative
size of the two order parameter fields near these two momenta defines
a 2D vector - hence the {Higgs}  (confinement/deconfinement) transition is in the
XY rather than Ising~\cite{MoonXu2012} universality class. {For any orientation of the 2D vector, the real space structure of the Higgs condensate is inhomogeneous.} (iv) The
critical theory $S_\phi + S_\psi + S_{\rm int}$ contains a
\textit{complex} boson $\phi$, two Dirac fermions $\psi$ and an
interaction term $S_{\rm int}$, which {in the XY case, however,} is RG-irrelevant~\cite{GroverSenthil2010}. (v)
According to Eq.~\eqref{eq:GTOT}, the confined phase inherits the small
Fermi surfaces of the deconfined OSM phase. By Oshikawa-Luttinger
theorem, this is only possible since lattice translational symmetry is
spontaneously broken in the confined semimetallic phase. (vi)
Nonetheless, the fermionic spectral weight is perfectly
translationally invariant, and for comparison to the metal plotted in
the large Brillouin zone in Fig.~\ref{fig:NonLinFS}~{\bf c}. (vii) The
quasiparticle residue of this spectral weight is momentum independent
and appears as $Z \sim \vert h^2 r/J^4 \vert^{2\beta}$, where $\beta =
0.35$ [(2+1)D XY].

\section{Implementation} 
\begin{figure}
\includegraphics[width = .48\textwidth]{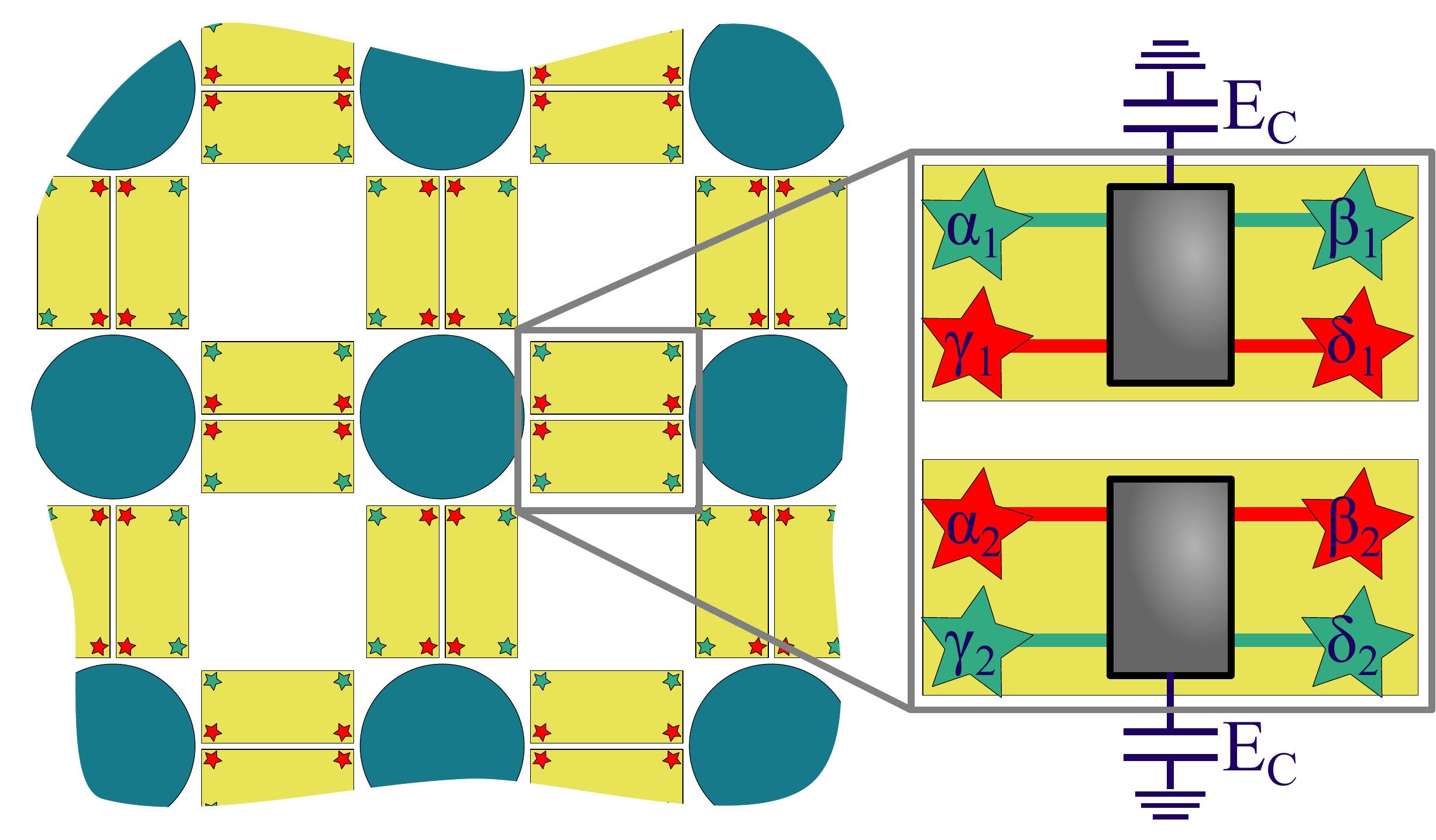}
\caption{A quantum emulator of Eq.~\eqref{eq:H0} is composed of an array of pairs of Majorana Cooper pair boxes (MCBs) (yellow rectangles in which Majorana modes are represented as stars), which encode the qubits on the links of Fig.~\ref{fig:model} {\bf a} and quantum dots (blue disks), which host the conduction electrons on the sites of Fig.~\ref{fig:model} {\bf a}. \textit{Inset:} Each MCB consists of two Kitaev chains which are coupled to a floating mesoscopic superconductor and capacitively coupled to the ground. If one out of two MCBs has occupied Majorana modes, a single charge can virtually hop from $\gamma_{1} \rightarrow \alpha_2$ and back from $\beta_2 \rightarrow \delta_1$, which lowers the energy (antiferromagnetic Ising superexchange).}
\label{fig:ToricCodeExperiment}
\end{figure}

We conclude by noting that our model can be implemented as a quantum circuit involving interpenetrating lattices of quantum dots and Majorana Cooper pair boxes (MCBs), Fig.~\ref{fig:ToricCodeExperiment} (see~App.~\ref{app:Emulator} for details). {This solid state proposal complements earlier proposals based on cold-atomic experimental setups~\cite{ZoharCirac2017}.}
On each MCB island, a large charging energy $E_C$ fixes the charge and thereby encodes a qubit~\cite{Fu2010,BeriCooper2012}, where the two degenerate quantum states, $\ket{\downarrow}$ ($\ket{\uparrow}$), have $N_0$ ($N_0-2$) particles in the condensate and empty (filled) pairs of Majorana modes. It has been proposed~\cite{TerhalDiVincenzo2012,LandauEgger2016}, that virtual hopping couples arrays of qubits to develop plaquette and star terms of the toric code. 

The interpenetrating array of quantum dots (represented as blue discs in Fig.~\ref{fig:ToricCodeExperiment}) 
contains one spinless electron per site and materializes the lattice sites (circles in Fig.~\ref{fig:model} {\bf a}). The logical qubits (squares in Fig.~\ref{fig:model} {\bf a}) are encoded in a pair of adjacent MCBs, in which easy-axis Ising superexchange selects a ground state Hilbert space spanned by $\ket{\uparrow_1, \downarrow_2},\ket{\downarrow_1, \uparrow_2}$. 

Virtual cyclic exchange of Majorana fermions around an empty plaquette generates $H_K$ in~Eq.~\eqref{eq:H0}, which due to the spatial separation of internal (red) and external (green) Marojana modes is unaffected by the quantum dots. Similarly, the star term derives from cyclic exchange of Majorana fermions around a quantum dot. Under appropriate tuning of microscopic parameters, it picks up an additional phase $\pi$ when the dot is occupied by electrons (i.e. $H_h$ in~Eq.~\eqref{eq:H0}). Finally, electron hopping between quantum dots occurs via a two step virtual process, hybridizing with the Majoranas at the red sites to produce the spin-dependent hopping term of strength $w$ in~Eq.~\eqref{eq:H0}. Experimental signatures of fermions in these artificial $\mathbb Z_2$ gauge theories, e.g. fermionic correlators, can be readily accessed by electronic coupling to the quantum dots. {We leave details of this and similar implementations, in particular the study of additional integrability breaking terms and  experimental signatures, to future studies}~\cite{KoenigInProgress}.

\section{Summary} In summary, inspired by parallels between
quantum information theory and correlated electron systems, by 
coupling Kitaev's
toric code to mobile fermions we have obtained a simple, solvable model for
fermions in the deconfined phase of a $\mathbb Z_2$ gauge theory {which can be simulated in an anologue quantum computer}. 
The phase diagram 
of this model contains a  
transition with 
abrupt changes in the Fermi surface topology between an orthogonal
metal and orthogonal semi-metal {through a sequence of symmetry breaking states}. We have {further} been able to
characterize {the Higgs transition} away from the integrable point using 
diagrammatic perturbation theory {and leave the quantum phase transitions between (orthogonal) semimetallic and metallic states at finite $J/h$ (blurry line in Fig.~\ref{fig:model}) to future studies.}

Although our model is too abstract for direct application 
to real materials, there are a number of interesting observations
that may be of experimental relevance. 
{For example, i}n our model the
small Fermi-surface phase 
displays
parallels with the phenomenology of the pseudogap phase in the
cuprates~\cite{KeimerZaanen2015}. {In light of such resemblances, we believe that the strategy of this work, i.e. condensing crucial emergent physical characteristics of quantum materials and analyzing them in a quantum information setting, will be a valuable scientific method for future research.}

\section{Acknowledgments} {We thank P.-Y.~Chang, Y.~Komijani,  N.~Perkins, M.~Scheurer, P.~Volkov for useful discussions.
Funding: PC and EJK were supported by the U.S. Department of Energy, Office of Basic Energy Sciences, under contract No. DE-FG02-99ER45790, AMT was supported by the U.S. Department of Energy, Office of Basic Energy Sciences, under contract No. {DE-}SC0012704.}

\appendix

\section{Relationship to Fermion-Ising-Higgs gauge theory}
\label{app:IsingHiggs}

In this section we explicitly relate the model of Gazit et al.\cite{GazitSachdev2019} to the fermionic toric code. We generalize the steps presented in~Ref.~\onlinecite{FradkinBook}.

\subsection{Fermion-Ising-Higgs gauge theory}

The model of Ref.~\onlinecite{GazitSachdev2019} contains $\bar{\sigma}^{x,y,z}_{\v b}$ Pauli matrices describing a $\mathbb Z_2$ gauge field which live on the bonds $\v b$ of a square lattice. It also contains $\tau^{x,y,z}_{\v r}$ Pauli matrices describing Higgs matter (slave spins) on the vertices $\v r$ of the same lattice and spinful fermions $f_{\v r, \alpha}$ living also on the vertices. The Hamiltonian is 
\begin{subequations}
\begin{equation}
\mathcal H = \mathcal H_{\mathbf Z_2} + \mathcal H_{\tau} + \mathcal H_f + \mathcal H_c + \mathcal H_U,
\end{equation}
where
\begin{eqnarray}
\mathcal H_{\mathbf Z_2} &=& - K \sum_{\square} \prod_{\v b \in \square} \bar \sigma_{\v b}^z - g \sum_{\v b} \bar \sigma_{\v b}^x ,\\
\mathcal H_{\tau} &=& - J  \sum_{\v b_{\v r, \v r'}} \bar \sigma_{\v b_{\v r, {\v r}'}}^z \tau_{\v r}^z \tau_{\v r'}^z - h \sum_{\v r} \tau_{\v r}^x ,\\
\mathcal H_{f} &=& - w \sum_{\v b_{\v r, \v r'}} \bar \sigma_{\v b_{\v r, \v r'}}^z f_{\v r,\alpha}^\dagger f_{\v r',\alpha},\\
\mathcal H_c &=& - t \sum_{\v b_{\v r, \v r'}} \tau_{\v r}^z f^\dagger_{\v r,\alpha} \tau_{\v r'}^z f_{\v r',\alpha} ,\\
\mathcal H_{U} &=& U \sum_{\v r} \left (\hat n_{\v r, \uparrow} - 1/2\right)\left (\hat n_{\v r, \downarrow} - 1/2\right).
\end{eqnarray}
\label{eq:HGAS}
\end{subequations}

The symmetries of this Hamiltonian are~\cite{GazitSachdev2019}
\begin{itemize}
	\item global SU(2) (spin): $f_{\v r,\alpha} \rightarrow U_{\alpha \beta} f_{\v r, \beta}$,
	\item at $\mu = 0$: global isospin SU(2) in particle-hole space,
	\item  local $\mathbf Z_2$ generated by the conserved charges $\bar Q_{\v r} = \underbrace{(-1)^{\hat n_{\v r}} \tau_{\v r}^x}_{\rm matter} \underbrace{\prod_{\v b \in \plus_{\v r}} \bar \sigma_{\v b}^x}_{\rm gauge\; field}$, where $\hat n_{\v r} = f^\dagger_{\v r,\alpha} f_{\v r, \alpha}$.
\end{itemize}

We highlight that part of this Hamiltonian, $\mathcal H_{\mathbb Z_2} + \mathcal H_{f}$ was studied before, e.g. in Refs.~\onlinecite{NandkishoreSenthil2012, GazitVishwanath2017}.

\subsection{Physical origin within $\mathbb Z_2$ slave spin theories}

{We briefly comment on the connection between this model and $\mathbb{Z}_2$ slave spin theory as introduced by R\"uegg, Huber and Sigrist~\cite{RueggSigrist2010}.}

{Consider a model of spinful fermions on a square lattice,
\begin{equation}
\mathcal H_{\rm Hubbard + \dots} = - t_0 \sum_{\v b_{\v r, \v r'}} c^\dagger_{\v r,\alpha}c_{\v r',\alpha} + \sum_{\v r} [U_0 \hat n_{\v r} (\hat n_{\v r}-1) - \mu \hat n_{\v r}] + \dots .\label{eq:HubbExt}
\end{equation}
The terms ``$\dots$'' represent additional non-local terms which are not further specified (see discussion below). In the case when the Hubbard $U_0$ is twice the chemical potential, the onsite problem contains two two-fold degenerate levels. These may be represented by a spin variable, using $\langle \tau^x_{\v r} \rangle = 1$ ($\langle \tau^x_{\v r} \rangle = -1$) for the singly occupied states at energy $-\mu$ (empty or doubly occupied states at energy $0$). In the slave spin formulation, we have thus fractionalized the conduction electrons
\begin{equation}
 c^\dagger_{\v r} = f^\dagger_{\v r} \tau^z_{\v r}.
 \end{equation}.
In this representation
\begin{equation}
\mathcal H_{\rm Hubbard + \dots} = - t_0 \sum_{\v b_{\v r, \v r'}} f^\dagger_{\v r,\alpha}\tau^z_{\v r}f_{\v r',\alpha}\tau^z_{\v r'} - U_0 \sum_{\v r} \tau_x + \dots,
\end{equation} 
using the onsite-constraint $(-1)^{\hat n_{\v r}} \tau^x_{\v r} = 1$. }
 
{Depending on the nature of the additional unspecified terms denoted `$\dots$' in the previous equation, a series of renormalization group steps may then lead to Eq.~\eqref{eq:HGAS} as an effective low-energy theory (with the more general gauge invariant constraint $\bar Q_{\v r} = 1$). By reversing the arguments, the physics discussed in this paper is applicable for those models which are in the basin of attraction of the ``fix-point Hamiltonian'' Eq.~\eqref{eq:HGAS}. In particular, note that $h$ in Eq.~\eqref{eq:HGAS} corresponds to the onsite repulsion of the extended Hubbard model Eq.~\eqref{eq:HubbExt}. }

\subsection{Formulation in terms of gauge invariant quantities.}

Since the total, local charge is conserved, we can impose Gauss' law on the physical Hilbert space
\begin{equation}
\bar Q_{\v r} \ket{\rm Phys} = \ket{\rm Phys} \quad \text{(``Gauss' law'')}.
\end{equation}

We readily see that gauge invariant quantities are
\begin{itemize}
\item ``c-electrons'': $c_{\v r, \alpha} = \tau_{\v r}^z f_{\v r, \alpha}$ ,
\item ``$\mathbb Z_2$ electric strings'' along a contour $\gamma_{\v r, \v r'}$ between $\v r, \v r'$: $ W^{(e)}_{\gamma_{\v r, \v r'}} = 	 \tau_{\v r}^z \left [\prod_{\v b \in \gamma_{\v r, \v r'}} \bar \sigma_{\v b}^z \right]\tau_{\v r'}^z 
\equiv \prod_{\v b \in \gamma_{\v r, \v r'}} \sigma_{\v b}^z$ .
\item Of course, $\bar\sigma_{\v b}^x \equiv  \sigma_{\v b}^x$ (``magnetic strings'') and $\hat n_{\v r} = f^\dagger_{\v r, \alpha} f_{\v r, \alpha} = c^\dagger_{\v r, \alpha} c_{\v r, \alpha}$ are trivially gauge invariant.
\end{itemize}

Here we have introduced Pauli matrices without a bar, $\sigma^{z}_{\v b} = \prod_{\v r \in \partial {\v b}} \tau^z_{\v r} \bar \sigma^z_{\v b}$ and $\sigma^{x}_{\v b} = \bar \sigma^x_{\v b}$, which clearly also have the appropriate commutation relations.
Except~\cite{FradkinBook} for special lines $J = 0$ or $g = 0$, all states of $\mathcal H_{\mathbb Z_2} + \mathcal H_\tau$ can be fully specified in the unitary gauge in which $\tau_{\v r}^z \ket{\rm Phys} = \ket{\rm Phys}$. (Indeed, within the physical subspace, where  $\tau_{\v r}^x \rightarrow (-1)^{\hat n_{\v r}} \prod_{\v b \in \plus_{\v r}} \sigma_{\v b}^x$, the Hamiltonian $\mathcal H$ preserves this gauge choice.) 
In short, having fixed the gauge $Q_{\v r} \ket{\rm Phys} = \ket{\rm Phys} = \tau_{\v r}^z \ket{\rm Phys}$ allows to express all gauge invariant quantities without resorting to $\tau$ operators  

\begin{subequations}
\begin{equation}
\mathcal H = \mathcal H_{\mathbf Z_2} + \mathcal H_{\tau} + \mathcal H_f + \mathcal H_c + \mathcal H_U,
\end{equation}
where
\begin{eqnarray}
\mathcal H_{\mathbf Z_2} &=& - K \sum_{\square} \prod_{\v b \in \square} \sigma_{\v b}^z - g \sum_{\v b} \sigma_{\v b}^x ,\\
\mathcal H_{\tau} &=& - J  \sum_{\v b} \sigma_{\v b}^z   - h \sum_{\v r}  \prod_{\v b \in \plus_{\v r}} {(-1)^{\hat n_{\v r}}} \sigma_{\v b}^x,\\
\mathcal H_{f} &=& - w \sum_{\v b_{\v r, \v r'}} \sigma_{\v b_{\v r, \v r'}}^z c_{\v r,\alpha}^\dagger c_{\v r',\alpha} ,\\
\mathcal H_c &=& - t \sum_{\v b_{\v r, \v r'}} c^\dagger_{\v r,\alpha}  c_{\v r',\alpha}, \\
\mathcal H_{U} &=& U \sum_{\v r} \left (\hat n_{\v r, \uparrow} - 1/2\right)\left (\hat n_{\v r, \downarrow} - 1/2\right).
\end{eqnarray}
\label{eq:H0FixedGauge}
\end{subequations}
This is a fermionic toric code, and at $g = J = t = U = 0$ the same as Eq.~\eqref{eq:H0} of the main text.

\section{Symmetries in the orthogonal semimetal phase}
\label{app:Symmetries}

\subsection{Explicit construction of the $\pi$-flux states}
\begin{figure}
\includegraphics[scale=.5]{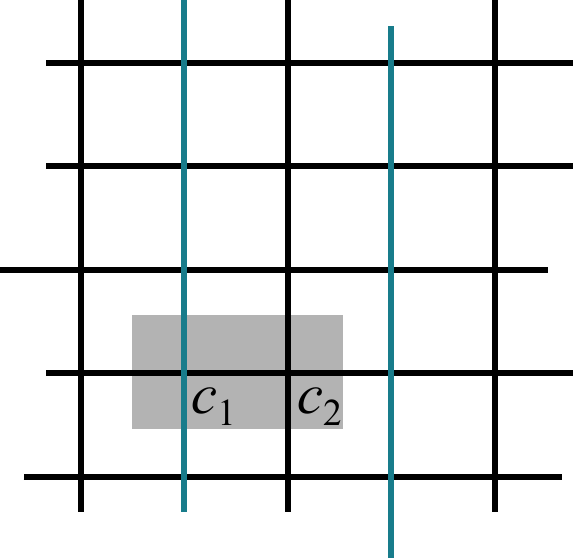}
\caption{Unit cell (shaded gray) and flux configuration prior to application of $Q_{\v r}$ operators in the OSM phase (green lines represent bonds with $\braket{\pi \vert \sigma^z_{\v b} \vert \pi}= -1$).}
\label{fig:PiFluxConfigs}
\end{figure}

Here, we solve the $\pi$ flux model in the gauge where $\langle  \sigma^z_{\v b} \rangle = (-1)^{b_x}$, i.e.~it is negative on every other vertical column but positive everywhere else (see Fig.~\ref{fig:PiFluxConfigs}). We choose a two atom unit cell of dimers along the $x$ direction and Fourier transform
\begin{eqnarray}
c_{\v x,1} &=& \int_{\rm small\; BZ}(dk) e^{i \v k \v r} c_{\v k,1}\\
c_{\v x,2} &=& \int_{\rm small\; BZ}(dk) e^{i \v k \v r + i k_x x} c_{\v k,2}
\end{eqnarray}  
(Note that $1,2$ labels do not correspond to sublattice labels $A,B$). The momentum space Hamiltonian is
\begin{equation}
H = -2 w \int_{\rm small\; BZ}(dk) c_{\v k}^\dagger [- \cos(k_y) \gamma_z + \cos(k_x) \gamma_x] c_{\v k}  \label{eq:PiFlux}
\end{equation}
[the small Brillouin zone (BZ) is $\v k \in (-\pi/2, \pi/2)\times(-\pi,\pi)$] which implies a dispersion 
\begin{equation}
\epsilon_\pi^{(\pm)}({\v k}) = \pm 2w \sqrt{\cos(k_x)^2 + \cos(k_y)^2}.
\end{equation}
Dirac nodes occur at $\vert k_x \vert = \vert k_y \vert = \pi/2$.

\subsection{Projective representation of translational symmetry}

In this section we explicitly demonstrate the projective representation of translational symmetry. For simplicity, we consider a system below half filling. For the gauge choice $ \ket{\pi}_\sigma$ discussed above, $\braket{\pi \vert \sigma^z_{\v b} \vert \pi}_\sigma = (-1)^{b_x}$, the Fermi surface is given by
\begin{eqnarray}
\ket{\rm FS}_c &=& \prod_{\v k \in \text{FS}}(c_{\v k,1}^\dagger,c_{\v k,2}^\dagger) \psi_{\v k}^{(-)} \ket{0} \notag \\
&=& \prod_{\v k \in \text{FS}} \sum_{\v x, \v x'} e^{i \v k (\v x - \v x')}(c_{\v x,1}^\dagger,c_{\v x,2}^\dagger) \psi_{\v x'}^{(-)} \ket{0}, \label{eq:state1}
\end{eqnarray}
where $\psi_{\v k}^{(-)}$ is the two-component eigenstate of $h(\v k) = -\cos(k_y)\gamma_z + \cos(k_x) \gamma_x$ with energy $\epsilon_{\pi}^{(-)}(\v k)$. 

We now consider a different gauge choice, $\ket{\pi'}_\sigma$, in which the columns of minus signs have been shifted by one lattice constant, $\braket{\pi' \vert \sigma^z_{\v b} \vert \pi'}_\sigma = (-1)^{b_x+1}$. It has a different Slater wave function, i.e.
\begin{eqnarray}
\ket{\rm FS'}_c &=& \prod_{\v k \in \text{FS}}(c_{\v k,1}^\dagger,c_{\v k,2}^\dagger) \tilde \psi_{\v k}^{(-)} \ket{0} \notag \\
&=& \prod_{\v k \in \text{FS}} \sum_{\v x, \v x'} e^{i \v k (\v x - \v x')}(c_{\v x,1}^\dagger,c_{\v x,2}^\dagger) \tilde \psi_{\v x'}^{(-)} \ket{0}.\label{eq:state2}
\end{eqnarray}
Since $\ket{\pi'}_\sigma$ induces a fermionic hopping Hamiltonian $\tilde h(\v p) = h(p_x, p_y + \pi)$, it follows that $\tilde \psi_{\v k}^{(-)} = \psi_{k_x, k_y + \pi}^{(-)}$.

We now demonstrate that $\ket{\pi'}_\sigma \ket{\rm FS'}_c = \prod_{\v r \in (\mathbb Z, 2 \mathbb Z)} Q_{\v r} \ket{\pi}_\sigma \ket{\rm FS}_c$ and hence $\prod_{\v r} \hat P_{\v r} \ket{\pi'}_\sigma \ket{\rm FS'}_c  = \prod_{\v r} \hat P_{\v r} \ket{\pi}_\sigma \ket{\rm FS}_c$ (i.e. the two seemingly different states Eq.~\eqref{eq:state1},\eqref{eq:state2} are the same in the deconfined phase: translational symmetry is restored). It is easy to see that the string of $Q_{\v r}$ operators translates the columnar gauge field pattern of negative bonds by one. The effect of the fermionic parity operator requires a little more explanations:
\begin{eqnarray}
&&\prod_{\v r \in (\mathbb Z, 2 \mathbb Z)}(-1)^{\hat n_{\v r}} \ket{\rm FS'}_c \notag \\
&&= \prod_{\v k \in \text{FS}} \sum_{\v x, \v x'} e^{i \v k (\v x - \v x')} (-1)^{y}(c_{\v x,1}^\dagger,c_{\v x,2}^\dagger) \tilde \psi_{\v x'}^{(-)} \ket{0} \notag\\
&&=\prod_{\v k \in \text{FS}} (c_{\v k,1}^\dagger,c_{\v k,2}^\dagger) \tilde \psi_{(k_x,k_y + \pi)}^{(-)} \ket{0} \notag \\
&&= \ket{\rm FS}_c.
\end{eqnarray} 
In the second line we have used that the spectrum (and thus the Fermi surface) is invariant under shifts of $\pi$ in $y$ direction.

\section{Luttinger's theorem}
\label{app:LuttingerTheorem}

\begin{figure}
\includegraphics[scale=.55]{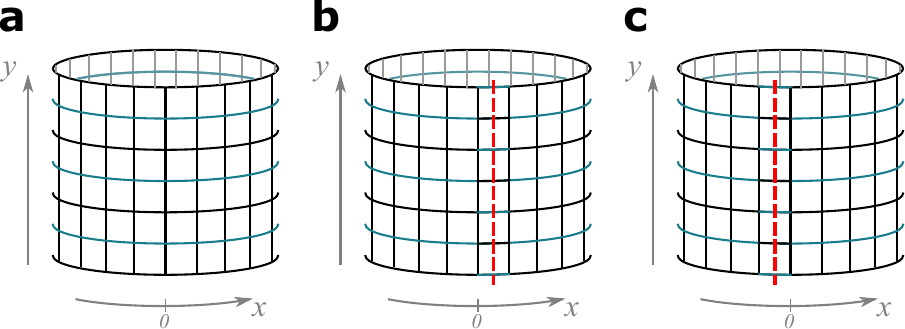}
\caption{{Flux insertion argument in the OSM phase. {\bf a} Representation of $\ket{\pi}_\sigma$ (analogous to Fig.~\ref{fig:PiFluxConfigs}) prior to the insertion of a $\pi$ flux. {\bf b} Configuration $\ket{\pi}_\sigma$ after insertion of a $\pi$ flux. {\bf c} Translation of the latter $\ket{\pi}_\sigma$.}}
\label{fig:Cylinder}
\end{figure}

{In this appendix we explain why the vanishing Fermi surface in the orthogonal semimetal at half filling is consistent with the Oshikawa-Luttinger-theorem~\cite{Oshikawa2000}. We follow arguments similar to those presented by Paramekanti and Vishwanath~\cite{ParamekantiVishwanath2004}.}

{To this end, we consider a our model, Eq.~\eqref{eq:H0}, on a finite cylinder as in Fig.~\ref{fig:Cylinder}.}

{We first remind the reader about the Oshikawa-Luttinger theorem for conventional Fermi liquids. We summarize the main physics and leave mathematical and technical details to the original literature. After the adiabatic insertion of a flux $2\pi$ through the hole of the cylinder the many-body Hamiltonian is gauge equivalent to the Hamiltonian prior to the insertion of the flux. Thus all eigenstates are the same. Still, by simple electrodynamics, the cylinder now spins due to the electromotive force and the total (angular) momentum of the system is $\Delta P_x = 2 \pi L_y n$ (where $n$ is the average particle number per site). In a Fermi liquid the quasiparticles excitations near the Fermi surface acquire a momentum because the Fermi distribution is shifted in $x$ direction. A simple integration by parts yields $\Delta P_x^{\rm FS} = L_y V_{\rm FS}/2\pi$, where $V_{\rm FS}$ is the volume enclosed by the Fermi surface. In  a Fermi liquid, the only momentum carrying particles are the quasiparticles and therefore $\Delta P_x = \Delta P_x^{\rm FS}\;  (\text{mod}\; 2 \pi L_y)$ and thus
\begin{equation}
\frac{V_{\rm FS}}{4\pi^2} = n \; (\text{mod}\; 1), \label{eq:LuttNormal}
\end{equation}
where the addition ``$\text{mod} \; 1$'' formally stems from crystalline momentum conservation modulo reciprocal lattice vectors and physically accounts for fully filled bands.}

{We consider a slightly different setup in the OSM phase. For the state $\ket{\pi}_\sigma$, we consider a gauge choice of the negative bonds as depicted in Fig.~\ref{fig:Cylinder} {\bf a} and thread a flux $\pi$ through the whole of the cylinder. This flux $\pi$ in the physical $U(1)$ field can be absorbed into a reconfiguration of $\sigma$ spins, Fig.~\ref{fig:Cylinder} {\bf b}. Despite having inserted only half a flux, it is thus evident that the Hamiltonian after flux insertion $H_{\pi}$ is related to $H_0$ of Eq.~\eqref{eq:H0} by a simple unitary transformation
\begin{equation}
H_{\pi} = W^{(m)} H_0 W^{(m)},
\end{equation} 
where $W^{(m)} = \prod_{\v b \in \mathcal C} \sigma^x_{\v b}$ is given by the magnetic string along the dashed contour presented in Fig.~\ref{fig:Cylinder}~{\bf b}. Obviously, $H_{\pi}$ and $H_0$ have the same spectrum, and the ground state has evolved from $\ket{GS_\pi}$ to $W^{(m)} \ket{GS_\pi}$.}

{We now demonstrate that, under the assumption $h >0$ and at half-filling, $W^{(m)} \ket{GS_\pi}$ carries the momentum transferred to the cylinder and at the same time the fermionic distribution function is unaffected. Thus, the momentum balance is accounted for by the deconfining gauge sector. Indeed, we can exploit that translation by one lattice site in $x$ direction can  is equivalent to application of $\prod_{\v r = (0,y)}\prod_{\v b\in \plus_{\v r}} \sigma_{\v b}^x = \prod_{\v r = (0,y)}(-1)^{\hat n_{\v r}} Q_{\v r}$ which acts solely on the gauge sector and leaves the fermions invariant. We can then use
\begin{equation}
\prod_{\v r = (0,y)}(-1)^{\hat n_{\v r}} Q_{\v r}W^{(m)} \ket{GS_\pi} = e^{i \pi L_y \langle \hat n_{\v r} \rangle} W^{(m)} \ket{GS_\pi}.
\end{equation}
Here, we used that $[Q_{\v r},W^{(m)}] = 0, Q_{\v r} \ket{GS_\pi} =  \ket{GS_\pi}$ and a homogeneous density. At half filling, the momentum balance between $\Delta P_x = \pi L_y n$ (from electrodynamics) and $\Delta P_x^{\ket{\pi}_\sigma} = \pi L_y \langle \hat n_{\v r} \rangle$ trivially reproduces $ n = \langle \hat n_{\v r} \rangle$ (valid only at half-filling $n = 1/2$) without resorting to the Fermi surface.}

{We now relax one of the two assumptions outlined above and consider a filling away from $1/2$. Then, in the OSM phase, the Luttinger-Oshikawa theorem implies a balance which is shifted by $1/2$ as compared to Eq.~\eqref{eq:LuttNormal} 
\begin{equation}
\frac{V_{\rm FS}}{4\pi^2}+ \frac{1}{2} = n \; (\text{mod}\; 1) . \label{eq:LuttZ2}
\end{equation}}

{Finally, we relax the second assumption for the derivation of the modified Luttinger theorem and comment on the case in which $h<0$ in which the ground state obeys $Q_{\v r} \ket{GS_{0/\pi}} = - \ket{GS_{0/\pi}}$. (Clearly, one may construct groundstats for this case analogously to Eq.~\eqref{eq:Solutions} in the main text.) For spinless fermions, the model Eq.~\eqref{eq:H0} at negative $h$ is related to the same model at positive   $h$ by a particle-hole transformation $c_{\v r} \leftrightarrow c^\dagger_{\v r}; \sigma^{x,z}_{\v b} \rightarrow -\sigma^{x,z}_{\v b}$. Therefore, all the conclusions obtained for Eq.~\eqref{eq:H0} at $h >0$ are applicable to $h<0$, as well.}

\section{Diagrammatic rules.}
\label{app:Diagram}

In this section we present the diagrammatic rules for imaginary time ordered, ground state correlators of fermionic operators $O_{\v r}(\tau) \in \lbrace c_{\v r}(\tau), c^\dagger_{\v r}(\tau) \rbrace$ and $\sigma^z_{\v b}(\tau)$ insertions
\begin{equation}
C(\lbrace \v r; \v b; \tau \rbrace) = -\braket{\text{GS} \vert \mathcal T[\prod_n O_{\v r_n}(\tau_n) \prod_m \sigma^z_{\v b_{m}}(\tau_m)] \vert \text{GS}}.
\end{equation}

\begin{enumerate}
\item Draw ``$\circ$'' for $c_{\v r}(\tau)$, ``$\bullet$'' for $c_{\v r}^\dagger(\tau)$, ``$\times$'' for $\sigma^z_{\v b}(\tau)$ at the corresponding position $\v r$, $\v b$ in real space. 
\item Only configurations with an even number of operators per site can be non-zero, $N_\circ + N_\bullet + N_\times\in 2 \mathbb N_0$ ({gauge field} operators $\sigma^z_{\v b}$ are associated to both adjacent sites $\v r \in \partial \v b$). 
\item Connect operators associated to a given site as follows. 
\begin{enumerate}
\item For two operators at times $\tau_1,\tau_2$, draw a wavy line which means $D(\tau_1, \tau_2) = e^{- 2h \vert \tau_1 - \tau_2\vert}$.  
\item For $2l>2$ operators at times $\tau_m$ ($m \in \lbrace {1}, \dots {2l}\rbrace$) encircle the operators which means 
$e^{- 2h \sum_{k = 1}^{2l} (-1)^k (\mathcal T\lbrace \tau_m\rbrace)_{k}},$ 
where $\mathcal T$ time orders the string of times $\lbrace \tau_m \rbrace$ in ascending order. 
\end{enumerate}
\item Evaluate all ``$\times$'' by $\braket{0 \vert \sigma^z_{\v b} \vert 0}_\sigma$ ($\braket{\pi \vert \sigma^z_{\v b} \vert \pi}_\sigma$) in the OM (OSM) phase. 
\item Connect all ``$\circ$'' and ``$\bullet$'' in all possible combinations according to the standard rules for fermionic diagrammatics with solid lines representing the ordinary Green's fuction $G_{\rm FS}(\v r_1, \v r_2; \tau_1, \tau_2) = - \braket{\text{FS}_{0/\pi} \vert \mathcal T [ c_{\v r_1}(\tau_1) c^\dagger_{\v r_2}(\tau_2)]\vert \text{FS}_{0/\pi}}$.
\end{enumerate}

Derivation of these rules.

\begin{enumerate}
\item We use $\ket {\text{GS}} = \prod_{\v r} \hat P_{\v r} \ket{\text{FS}}_c\ket{0/\pi}_\sigma$, and pass the projectors over the string of operators to find that $\braket{\text{GS} \vert \mathcal T[O_{\v r_n}(\tau_n) \dots O_{\v r_1}(\tau_1) \sigma^z_{\v b_m'}(\tau_m') \dots \sigma^z_{\v b_1'}(\tau_1')] \vert \text{GS}}$ vanishes, unless an even number of operators is associated to each site (each $\sigma^z_{\v b}$ is associated to both adjacent sites $\v r\in \partial \v b$). Therefore
\begin{itemize}
	\item all fermionic operators are connected by ``electric'' strings of $\sigma^z$ (includes the possibility of two fermions on the same position, i.e. string of zero extension)
	\item all electric strings are either closed or end in fermionic operators.
\end{itemize}

 [We used $\hat P_{\v r} = [1+ Q_{\v r}]/2$ and we will repeatedly use $Q_{\v r}  O_{\v r'} = (-1)^{\delta_{\v r, \v r'}} O_{\v r'} Q_{\v r} $ and $Q_{\v r} \sigma^z_{\v b} = (-1)^{\delta_{\v r \in \partial \v b}} \sigma^z_{\v b}Q_{\v r}  $.] 

\item The interaction picture representation of the fermionic operators $O_{\v r} \in \lbrace c_{\v r}, c^\dagger_{\v r}\rbrace$ is $O_{\v r}(\tau) = e^{H_0\tau} O_{\v r} e^{- H_0 \tau} = \bar O_{\v r}(\tau) e^{2 h Q_{\v r} \tau}$, where $\bar O (\tau) = e^{H_c\tau}O_{\v r} e^{-H_c\tau}$. Similarly, the interaction picture representation of $\sigma^z_{\v b}(\tau) = \sigma^z_{\v b}  e^{2 h \sum_{\v r \in \partial \v b} Q_{\v r} \tau}$.
\item We (i) explicitly time order the string of operators in $C(\lbrace \v r, \tau \rbrace)$, (ii) use the representation $O_{\v r}(\tau) =  \bar O_{\v r}(\tau) e^{2 h Q_{\v r} \tau}$, $\sigma^z_{\v b}(\tau) = \sigma^z_{\v b}  e^{2 h \sum_{\v r \in \partial \v b} Q_{\v r} \tau}$ that we just derived, (iii) then pass all $e^{2 h Q_{\v r} \tau}$ to the right of all $\bar O_{\v r} (\tau), \sigma^z_{\v b}$ using again $Q_{\v r} \bar O_{\v r'} (\tau) = (-1)^{\delta_{\v r, \v r'}} \bar O_{\v r'} (\tau) Q_{\v r} $, $Q_{\v r} \sigma^z_{\v b} = (-1)^{\delta_{\v r \in \partial \v b}} \sigma^z_{\v b}Q_{\v r}  $ and (iv) finally use $Q_{\v r} \ket {\text{GS}} = \ket {\text{GS}}$ to obtain Feynman rule No. 3 (e.g. exponentials of the kind $e^{-2h\vert \tau - \tau' \vert}$ represented by wavy lines). 
\item At this point the correlator has been evaluated to be $C(\lbrace \v r,  \tau\rbrace)=-\braket{\text{GS} \vert \mathcal T[\bar O_{\v r_n}(\tau_n) \dots \bar O_{\v r_1}(\tau_1) \sigma^z_{\v b_m'} \dots\sigma^z_{\v b_1'}]\vert \text{GS}} \times \text{\textit{(exponentials represented by wavy lines)}}$. The only gauge field (= spin $\sigma$) dependence in the operators is now inside $H_c$ entering $\bar O(\tau)$ and in the strings of $\sigma^z$. We can thus replace all gauge fields by the ground state (e.g. all up in zero flux) configuration $\sigma^z \rightarrow \langle \sigma^z \rangle_{\ket{0}/\ket{\pi}}$, $H_c \rightarrow H_{0/\pi}$ 
\item We have brought the correlator to  the form $C(\lbrace \v r, \tau \rbrace)=-\braket{\text{FS} \vert \mathcal T[\bar O_{\v r_n}(\tau_n) \dots \bar O_{\v r_1}(\tau_1)] \vert \text{FS}} \times \text{\textit{(exponentials represented by wavy lines)}}$. Now we can use the standard Wick's theorem for fermions, this is Feynman rule No. 4.
\end{enumerate}

 Analogous rules hold upon Wick-rotation to real time $\tau \rightarrow i t$, $\vert \tau_1 - \tau_2 \vert \rightarrow i \vert t_1 - t_2 \vert$.

\section{{Higgs} transition}
\label{app:ConfDeconfTransition}

In this section we present details on the transition from deconfined to confined phases and a derivation of the effective field theory.

\subsection{Zero-flux case (starting point: orthogonal metal)}

\subsubsection{Propagator of ``e'' particles.}

The propagator of $\mathbf{D}(\v r_f, \v r_i; \tau_f, \tau_i)$ entering Eq.~\eqref{eq:GTOT} of the main text is defined according to diagram Fig.~\ref{fig:Perturbations} {\bf e} by 
\begin{widetext}
\begin{eqnarray}
\mathbf{D}(\v r_f, \v r_i; \tau_f, \tau_i) &=& D(\tau_f, \tau_i) \delta_{\v r_f, \v r_i} +  J \int d\tau D(\tau_f, \tau) D(\tau, \tau_i) \delta_{< \v r_f, \v r_i >} \notag \\ &+& J^2 \sum_{\substack{\v b, \v b' \text{s.th.} \notag \\ \v r_f \in \partial \v b, \v r_i \in \partial \v b'}}\int d\tau d \tau' D(\tau_f, \tau)  \langle \sigma^z_{\v b} (\tau) \sigma^z_{\v b'} (\tau') \rangle D(\tau', \tau_i) \left [1 - \delta_{< \v r_f, \v r_i >} - \delta_{\v r_f, \v r_i }\right].
\end{eqnarray}
{Here, $\delta_{< \v r_f, \v r_i >}=1$ for nearest neighbors, it vanishes otherwise.}

The resummation of non-intersecting strings of $\times$ insertions, Fig.~\ref{fig:Perturbations}~{\bf e} of the main text, is given by 
\begin{eqnarray}
\mathbf D(\v r_f, \v r_i; i \omega) &=&  D( i \omega) \delta_{\v r_f, \v r_i} + J \sum_{\v r \text {n.N. of } \v r_f} \braket{0 \vert \sigma^z_{\langle \v r, \v r_f \rangle } \vert 0}_\sigma \mathbf D(\v r, \v r_i; i \omega) D(i \omega). \label{eq:ResumD}
\end{eqnarray}
\end{widetext}

In momentum space this implies $\mathbf D(\v q, i \omega) = D(i\omega) + 2 J [\cos(q_x) + \cos(q_y)] D(i \omega) \mathbf D(\v q, i\omega)$ which immediately implies Eq.~\eqref{eq:DResum} of the main text. The inclusion of fermionic hopping, see Fig.~3~{\bf f}, implies $\mathbf D_t(\v q, i \omega) = \mathbf D(\v q, i\omega) + 2 \bar t [\cos(q_x) + \cos(q_y)] \mathbf D(\v q, i \omega) \mathbf D_t(\v q, i\omega)$
and thus
\begin{equation}
\mathbf D_t(\v q, i \omega) = \frac{4h}{\omega^2 + 4h (h - 2 (J + \bar t) [\cos(q_x) + \cos(q_y)])}.
\end{equation} 

\subsubsection{Self-interaction of electric strings.}

To obtain the mean field expectation value of $Z$ we first analyze non-linearities in $\v D(\v q, i\omega)$, which we extract from the connected part of $4-$point correlations functions.

In the absence of $t$, the non-linearity stems from Fig.~\ref{fig:NonLinFS}~{\bf a} of the main text, i.e.
\begin{eqnarray}
V(\lbrace \v r,\tau\rbrace ) &=& J^4 \big \lbrace \sum_{\v r} \prod_{n = 1}^4\delta_{<\v r_n, \v r >}
 \notag\\
& &\times  e^{- 2h [(\mathcal T\lbrace \tau\rbrace)_{4}-(\mathcal T\lbrace \tau\rbrace)_{3} + (\mathcal T\lbrace \tau\rbrace)_{2} -  (\mathcal T\lbrace \tau\rbrace)_{1}]} \notag \\
  &&- D(\tau_1,\tau_2) D(\tau_3, \tau_4) - D(\tau_1, \tau_3) D(\tau_2, \tau_4)\notag \\
  && - D(\tau_1, \tau_4) D(\tau_3, \tau_2)\big \rbrace. \label{eq:NonlinearitySimple}
\end{eqnarray}
We remind the reader, that for more than two wavy lines on a given site the time dependence of interaction is rather complicated in view of Feynman rule No. 3. Since we only keep the connected part of the 4-point correlator, we have subtracted the contributions which correspond to two ``e''-particle propagators running through each other without interaction.

For the derivation of the continuum field theory, we evaluate Eq.~\eqref{eq:NonlinearitySimple} at zero incident frequencies and obtain ($\beta = 1/T$ is the inverse temperatue, viz. the IR cut-off)
\begin{eqnarray}
\int \prod_n d \tau_n V(\lbrace \v r, \tau \rbrace) &=& \big \lbrace \frac{J^4}{h^2} \left ( 3 \beta^2 - 6 \beta/h + 9/(2h^2) \right) \notag \\
&&- 3\frac{J^4}{h^2} \left ( \beta - 1/(2h) \right)^2 \big \rbrace \sum_{\v r}\delta_{<\v r_n, \v r >} \notag \\
&\simeq&  - J^4 \frac{3 \beta}{h^3} \sum_{\v r}\delta_{<\v r_n, \v r >}. \label{eq:IA}
\end{eqnarray}
The bare value of the coefficent $\lambda \sim a^2 J^4/h$ follows from comparison of Eq.~\eqref{eq:IA} and \eqref{eq:EffFieldTheoryOM} of the main text. We remind that in Eq.~\eqref{eq:EffFieldTheoryOM}, the field $\phi$ was rescaled by $\sqrt{4h a^2}$ to compensate the numerator of the Green's function and to obtain the continuum limit.

\subsubsection{Interactions between fermions and strings.}

In the presence of $t$ perturbations, there are new operator insertions which would imply mutual impact of fermionic excitations and string in Fig.~3~{\bf e}. The inclusion of nearest neighbor hopping $t$ yields a local in time interaction
\begin{equation}
V( \v r_c, \v r_{c^\dagger},\v r_{D_1},\v r_{D_2})  = t \left [\delta_{\v r_c, \v r_{\rm D_1}} \delta_{\v r_{c^\dagger}, \v r_{\rm D_2}} + 1 \leftrightarrow 2 \right ] \delta_{\langle \v r_c, \v r_c^\dagger \rangle}.
\end{equation}

Upon rescaling of fields (fermions are rescaled by $a$) this yields a coupling constant $g \sim h t a^2 (\cos(\hat p_x) + \cos(\hat p_y))$ in the long-wavelength limit of $\phi$ fields.

\subsection{$\pi$-flux case (starting point: orthogonal semimetal)}

\subsubsection{Propagator of ``e'' particles}
In the $\pi$ flux case the, the resummation
\begin{align}
\mathbf D(\v r_f, \v r_i; i \omega) &=  D( i \omega) \delta_{\v r_f, \v r_i} \notag \\
+ J &\sum_{\substack{\v r \text{ n.N.} \\ \text{of } \v r_f}} \braket{\pi \vert \sigma^z_{\langle \v r, \v r' \rangle } \vert \pi}_\sigma \mathbf D(\v r, \v r_i; i \omega) D(i \omega) \label{eq:ResumDPi}
\end{align} 
contains the expectation value of $\sigma^z$ with respect to the $\pi$-flux state. Thus, we have to consider a matrix Green's function 
\begin{equation}
\underline{\mathbf D} (\v q, i\omega) = \left (\begin{array}{cc}
{\mathbf D}_{11} (\v q, i\omega) & {\mathbf D}_{12} (\v q, i\omega) \\ 
{\mathbf D}_{21} (\v q, i\omega) & {\mathbf D}_{22} (\v q, i\omega)
\end{array} \right )
\end{equation}
where the unit cell is as in Fig.~\ref{fig:PiFluxConfigs}. The Fourier transform of Eq.~\eqref{eq:ResumDPi} in matrix notation is thus
\begin{align}
\underline{\mathbf D} (\v q, i\omega) &= D(i\omega) \mathbf 1_{\gamma} \notag \\
&+ 2 J D(i \omega) [\cos(q_x) \gamma_x - \cos(q_y) \gamma_z] \underline{\mathbf D} (\v q, i\omega)
\end{align}

and therefore
\begin{equation}
\underline{\mathbf D} (\v q, i\omega) = 4 h\lbrace\omega^2 + 4 h (h - 2 J[\cos(q_x) \gamma_x - \cos(q_y) \gamma_z])\rbrace^{-1}.
\end{equation}

This propagator has two transitions happening simultaneously: one at $\v q  = (0,0)$ and one at $\v q = (0,\pi)$ (recall that $\v q \in (-\pi/2,\pi/2) \times (-\pi,\pi)$). 
The inclusion of $t$ implies a shift $J \rightarrow J + \bar t$ where $\bar t = 2t G_{\rm FS}(\v r, \v r'; \tau,\tau)$ and $\v r, \v r'$ are nearest neighbours.

\subsubsection{Self-interaction of electric strings}

The $\phi^4$ theory for the $\pi$ flux can be regarded as  [$\vec \phi = (\phi_1, \phi_2)$ lives on the two basis sites of the unit cell, Fig.~\ref{fig:PiFluxConfigs}]

\begin{widetext}
\begin{equation}
S [\vec \phi] = \int d \tau (dq) \vec \phi(-\v q,\tau) \frac{[-\partial_\tau^2 + 4 h (h - 2 J[\cos(q_x) \gamma_x - \cos(q_y) \gamma_z])]}{2} \vec \phi(\v q, \tau) + \frac{ \lambda}{4!} [\phi_1(\v x, \tau)^4 + \phi_2(\v x, \tau)^4].
\end{equation}
The locality of interactions in real space of Fig.~4~{\bf b} implies the $\phi_1^4 + \phi_2^4$ form of interactions and $ \lambda \sim a^2 J^4/h$. To derive the critical theory, we diagonalize the quadratic term, the bottom of the band is near $q_x = 0$ and gapped, so that it is sufficient to only consider the wave functions of the lower band. Then
\begin{equation}
S [\phi_0] = \int d \tau (dq)  \phi_-(-\v q,\tau) \frac{[-\partial_\tau^2 + 4 h (h - 2 J[\sqrt{\cos(q_x)^2 + \cos(q_y)^2 }])]}{2} \phi_-(\v q, \tau) + \frac{ \lambda}{4!} [\phi_-(\v x, \tau)^4].
\end{equation}
\end{widetext}
A factor of order unity has been absorbed into $ \lambda$.
In a subsequent step we expand near the position of the minima of the $\phi_-$ field: $\phi_-(\v x,\tau) \simeq \phi_0(\v x, \tau) + \phi_\pi(\v x, \tau) e^{i \pi y} $, where both $\phi_{0}$ and $\phi_\pi$ are slow fields. We group them into a complex field $\phi = \phi_0 + i \phi_\pi$ and obtain the effective theory (the relative weight of $\phi_0^4 + \phi_\pi^4$ and $\phi_0^2 \phi_\pi^2$ follows from momentum conservation)
\begin{equation}
S [\phi] = \int d \tau d^2 x\; \bar \phi {[-\partial_\tau^2- v^2 \nabla^2 + 4 h (h - 2 J\sqrt{2}))]} \phi + \frac{\lambda}{2} \vert \phi \vert^4.
\end{equation}
Again, we absorbed a factor of order one into $\lambda$.

{We now discuss the real-space pattern of the condesned Higgs field $\phi$ assuming a parametrization $\phi_0 \sim \cos(\varphi),\phi_\pi \sim \sin(\varphi)$, where $\varphi$ is the $XY$ angle which orders at the transtion. Then, in a given unit cell
\begin{equation}
\left (\begin{array}{c}
\phi_1 \\ 
\phi_2
\end{array}  \right) \propto \cos(\varphi) \left (\begin{array}{c}
\sqrt{2 - \sqrt{2}} \\ 
\sqrt{2 + \sqrt{2}}
\end{array}  \right)+(-1)^{y}\sin(\varphi) \left (\begin{array}{c}
\sqrt{2 + \sqrt{2}} \\ 
\sqrt{2 - \sqrt{2}}
\end{array}  \right),
\end{equation} 
where the vector structure follows from the eigenstates of $\cos(q_x) \gamma_x - \cos(q_y) \gamma_z$ at $\v q= 0, \v q= (0,\pi)$. Translational invariance in $x$ direction implies $\phi_1 = \phi_2$ and thus
\begin{equation}
\cos(\varphi) - (-1)^y\sin(\varphi)=0,
\end{equation}
while additionally imposing translational symmetry in $y$ direction implies that the equation shall be valid for any row $y$. This is impossible, therefore the Higgs field $(\phi_1, \phi_2)(\v x)$ condenses always in an inhomogeneous state breaking the crystalline symmetries of the model. 
}
\subsubsection{Interaction between fermions and strings.}

To obtain the effective interaction of fermions and strings, we consider the microscopic interaction, Fig.~\ref{fig:NonLinFS} of the main text,
\begin{eqnarray}
H_{\rm int} &=& t \sum_{\v r}[ c_1^\dagger (\v r) c_2(\v r) \phi_1(\v r) \phi_2(\v r) \notag \\
&+&  c_2^\dagger (\v r) c_1(\v r + 2 \hat e_x) \phi_2(\v r) \phi_1(\v r+ 2 \hat e_x) \notag \\
&+& c_1^\dagger(\v r) c_1(\v r+ \hat e_y) \phi_1(\v r) \phi_1(\v r + \hat e_y) \notag \\
&+& c_2^\dagger(\v r) c_2(\v r+ \hat e_y) \phi_2(\v r) \phi_2(\v r + \hat e_y) ] + h.c. \label{eq:HFString}
\end{eqnarray}
We consider only the coupling to the critical modes, i.e. 
\begin{eqnarray}
\vec \phi(\v x) &\simeq& \phi_0(\v x) \frac{1}{\sqrt{4 - 2\sqrt{2}}}\left (\begin{array}{c}
-1 + \sqrt{2} \\ 
1
\end{array} \right ) \notag \\
&&+ (-1)^y \phi_\pi(\v x) \frac{1}{\sqrt{4 - 2\sqrt{2}}}\left (\begin{array}{c} 1\\
-1 + \sqrt{2}
\end{array} \right ).
\end{eqnarray}

Thus we obtain
\begin{eqnarray}
\phi_1(\v r) \phi_2(\v r)&\simeq& \frac{\phi_0^2 + \phi_\pi^2}{2 \sqrt{2}} + (-1)^y\phi_0 \phi_\pi , \\
\phi_2(\v r) \phi_1(\v r + 2 \hat e_x)&\simeq& \frac{\phi_0^2 + \phi_\pi^2}{2 \sqrt{2}} + (-1)^y\phi_0 \phi_\pi , \\
\phi_1(\v r) \phi_1(\v r + \hat e_y) & \simeq &
- \frac{\phi_0^2 + \phi_\pi^2}{2 \sqrt{2}} + \frac{\phi_0^2 - \phi_\pi^2}{2},\\
\phi_2(\v r) \phi_2(\v r + \hat e_y) & \simeq &
\frac{\phi_0^2 + \phi_\pi^2}{2 \sqrt{2}} + \frac{\phi_0^2 - \phi_\pi^2}{2} .
\end{eqnarray}

We can now study the low energy theory near the Dirac nodes, using the spinor $\psi = (\psi_{1, \pi/2},\psi_{2, \pi/2}, \psi_{2, -\pi/2},\psi_{2, -\pi/2})$, where $1,2$ denotes the sublattice position and $\pm \pi/2$ the $y$ coordinate in the Brillouin zone $(0,\pi) \times(-\pi,\pi)$). The effective kinetic Hamiltonian near the Dirac nodes takes the form 
\begin{equation}
h(\v p) = -w [p_x \gamma_x \mathbf 1_\tau - p_y \gamma_y \tau_z] .
\end{equation}
Interactions between fermions and critical electric strings are given by
\begin{eqnarray}
S_{\rm int} &\sim& a^2 t h \int d \tau d^2x \Big \lbrace \vert \phi \vert^2 \bar \psi \left [\frac{h(-i \nabla)}{w} \right] \psi \notag \\ 
&&+ 2 \sqrt{2} \phi_0 \phi_\pi \bar \psi (-i \nabla_x) \tau_x \gamma_x \psi\notag \\
& &- \sqrt{2} [\phi_0^2 - \phi_\pi^2] \bar \psi (-i \nabla_y) \psi \Big \rbrace.
\end{eqnarray}

\section{Analogue Quantum Computer using Majorana Cooper pair boxes.}
\label{app:Emulator}

In this section we provide further details about the implementation of fermionic $\mathbb Z_2$ gauge theories using analogue quantum computers based on Majorana Cooper pair boxes. 

\begin{figure}
\includegraphics[scale=.22]{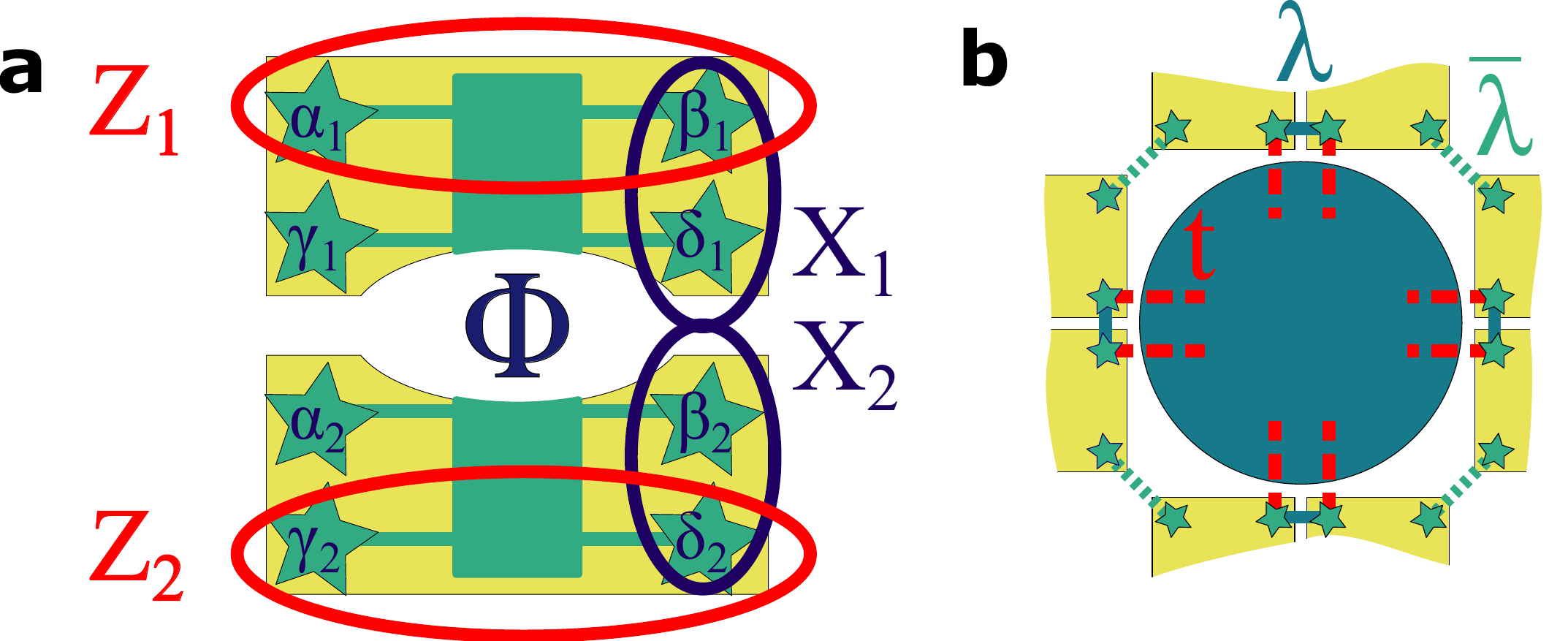}
\caption{Details on the quantum emulator presented in Fig.~\ref{fig:ToricCodeExperiment} of the main text. {\bf a} Each block emulating the toric code degrees of freedom consists of two MCBs which are threaded by a flux. When $E_C$ is the largest scale, each MCB is a two-level system in which $Z$ and $X$ gates (Pauli matrices) can be defined as bilinears of Majoranas, see Eq.~\eqref{eq:DefZXGates}. Virtual superexchange induces a pseudomagnetic Ising coupling between adjacent MCBs. Note that our convention of labelling Majoranas in vertical boxes is the counterclockwise 90$^\circ$ rotation of panel {\bf a}, whence $Z$ gates are always along the elongated sides of the rectangle. {\bf b} The Majorana modes and the fermions on the quantum dot are coupled by hopping matrix elements of amplitude $\lambda, \bar \lambda, t$. }
\label{fig:QuantumEmulator}
\end{figure}

The basic building block, Fig.~\ref{fig:ToricCodeExperiment}~{\bf b} and~\ref{fig:QuantumEmulator}~{\bf a}, to emulate the toric code sector of our model is a pair of Majorana Cooper boxes (MCB). A standard setup for each such MCB (here ``1'' for top and ``2'' for bottom) consists of two Kitaev wires which are contacted to a mesoscopic superconductor. The whole box is capacitively coupled to the ground by $E_C (\hat N_{1,2} - N_{0})^2$ (eigenvalues of $\hat N_{1,2}$ are integers, in this convention the condensate can absorb two units). The parity of the total number of electrons on the island fixes the parity of the Majorana sector, e.g. $\alpha \beta \gamma \delta = -1$. In this subspace we introduce Pauli-Matrices, see Ref.~\onlinecite{TerhalDiVincenzo2012} and Fig.~\ref{fig:QuantumEmulator} {\bf a},
\begin{subequations}
\begin{eqnarray}
Z &=& i \alpha \beta \doteq i  \gamma \delta,\\
X &=& i \alpha \gamma \doteq i \delta \beta.
\end{eqnarray}
\label{eq:DefZXGates}
\end{subequations}
We used the projector onto the low-energy subspace at ``$\doteq$''. 

Hopping matrix elements, see Fig.~\ref{fig:QuantumEmulator} {\bf b},
between the Majorana end modes at the wires and electrons in the dots are
\begin{align}
H_\lambda &=  \lambda_{\gamma \alpha} \gamma_l \alpha_r e^{ i \hat \phi_l/2 - i \hat \phi_r/2} + H.c. \text{ (top pair of Fig.~\ref{fig:QuantumEmulator} {\bf b})},\\
H_{\bar \lambda} &=  \bar \lambda \gamma_{rb} \delta_{br} e^{ i \hat \phi_{rb}/2 - i \hat \phi_{br}/2} \text{ (bottom right of Fig.~\ref{fig:QuantumEmulator} {\bf b})},\\
H_{t} &=  t_{c^\dagger \alpha} c^\dagger \alpha e^{- i \hat \phi_r/2} + H.c. \text{ (top right in Fig.~\ref{fig:QuantumEmulator} {\bf b})}.
\end{align}

We will drop the subscript of matrix elements and asssume $\bar \lambda, \lambda$ and $t$ to have the same amplitude everywhere as displayed in Fig.~\ref{fig:QuantumEmulator} {\bf b}. 

A  pair of adjacent MCBs is coupled by superexchange processes $H_{\rm SX} = 2\vert \lambda\vert^2 \cos(\Phi) Z_1 Z_2/E_C$, such that for enclosed flux $\Phi < \pi/2$ the effective logical qubit states are $\ket{\uparrow_1, \downarrow_2},\ket{\downarrow_1, \uparrow_2}$ and $\sigma^z = Z_1/2 -Z_2/2$, $\sigma^x = X_1X_2$. We project the toric code sector onto the logical, low-energy subspace, and obtain~Eq.~\eqref{eq:H0} of the main text using perturbation theory~\cite{TerhalDiVincenzo2012,LandauEgger2016} 

\begin{eqnarray}
w&\sim& - \vert t \vert^2 \sin(\Phi/2)/E_C,\\ 
K &\sim& -\vert \bar \lambda \vert^4\cos(\Phi_\square)/E_C^3,\\ 
h &\sim&  \mathcal C_1\vert t \vert^2{\vert \bar \lambda \vert^4}\vert \lambda \vert^3/{E_C^8} + \mathcal C_2 {\vert t \vert^6}{\vert \bar \lambda \vert^4}/{\vert \lambda \vert^3}{E_C^6}
\end{eqnarray} 
(with $\Phi$ dependent parameters $\mathcal C_{1,2}$). 
The impact of the star term without fermionic parity, which has a coupling 
\begin{equation}
h' = \frac{\vert \bar \lambda \vert^4}{E_C^4}  \left(\mathcal C_3 \frac{\vert \lambda \vert^4}{E_C^3} - \mathcal C_4 \frac{\vert t \vert^4}{E_C^3}- \mathcal C_5 \frac{\vert t^2 \bar \lambda \vert^4}{\vert \lambda \vert^6 E_C}\right).
\end{equation}
can be mitigated by appropriate tuning of $\vert t /\lambda \vert$ since the dimensionless parameters $\mathcal C_{1,2,3,4,5}$ are positive for small flux $\Phi$ through a pair of MCBs.

\section{OSM - OM transition: Flux configurations}
\label{app:OSMOMTransition}

In this section we present numerical details on the flux configurations underlying the transition from OSM to OM, as discussed in Sec.~\ref{sec:OSMOMTransition} of the main text.

In order to illustrate the physics at the small to large Fermi surface transition, we numerically diagonalized the Hamiltonians associated to a variety of flux configurations with average flux $\Phi = k \pi/8$, $k = 0, \dots, 8$ and determined fermionic energy $E_{c,\Phi}$ and particle density (=filling) $n$ at temperature $T = w/100$, for system size $40 \times 40$, and chemical potential $E_F \in [-w,w]$ (step size $\Delta E_F = w/20$). Clearly, at $k = 0,8$ simple analytical calculations could be used to check the numerical results. We subsequently fitted the numerical data to a symmetric 8-th order polynomial and thereby obtained an approximate function $E_{c,\Phi}(n)$ for each of the 38 configurations. The chosen range of chemical potentials allows reliable fits within a density $n \in [0.25,0.75]$. Finally, we determined the flux associated to the minimal total energy $E_{\rm tot}(K,\rho) =\min_{{\Phi,\text{config's}}}[-K \Phi + 2 E_{c,\Phi}(\rho)]$ and plotted it as a density plot in Fig.~1 {\bf b}. We have explicitly checked that particle-hole-symmetry is present in the phase diagram and therefore only plot $n >1/2$.

In Figs.~\ref{fig:ConfigsPi8}-\ref{fig:Configs7Pi8} we summarize the considered flux configurations, along with the associated numerical data and fits of $E_{c,\Phi}(n)$. In the schematic pictures of the lattice, blue bonds represent hopping matrix elements $-w$ and red dots a $\pi$ flux threading a plaquette. We considered three types of 16-site unit cells (shaded gray in the figures), thus the starting point of the numerics are the $16\times 16$ momentum space matrix Hamiltonians. 

\begin{widetext}

\begin{figure}
\begin{minipage}{\textwidth}
\includegraphics[scale=.4]{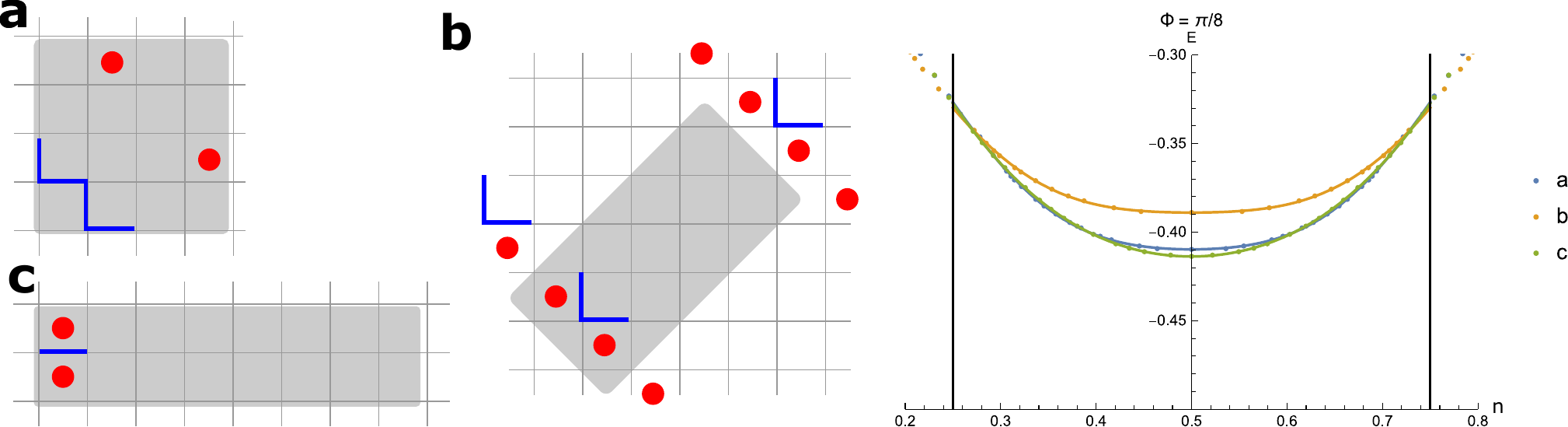}
\caption{Configurations of average flux $\Phi = \pi/8$ and associated fermionic ground state energies as a function of filling.}
\label{fig:ConfigsPi8}
\end{minipage}

\begin{minipage}{\textwidth}
\includegraphics[scale=.4]{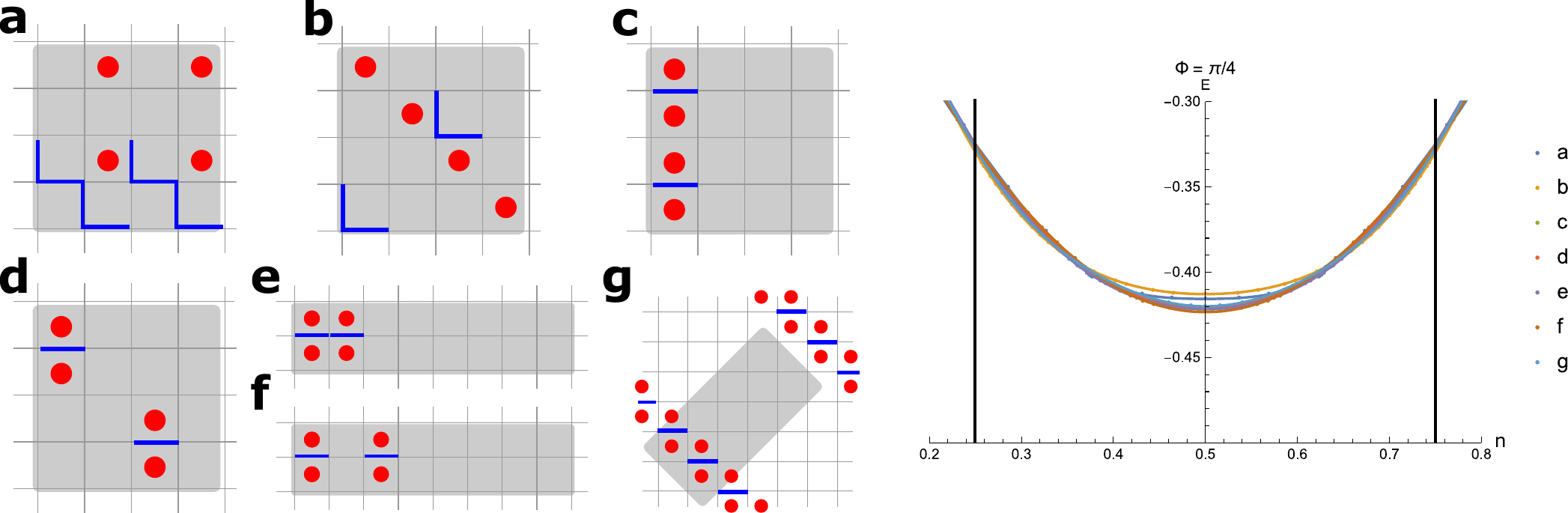}
\caption{Configurations of average flux $\Phi = \pi/4$ and associated fermionic ground state energies as a function of filling.}
\label{fig:ConfigsPi4}
\end{minipage}

\begin{minipage}{\textwidth}
\includegraphics[scale=.4]{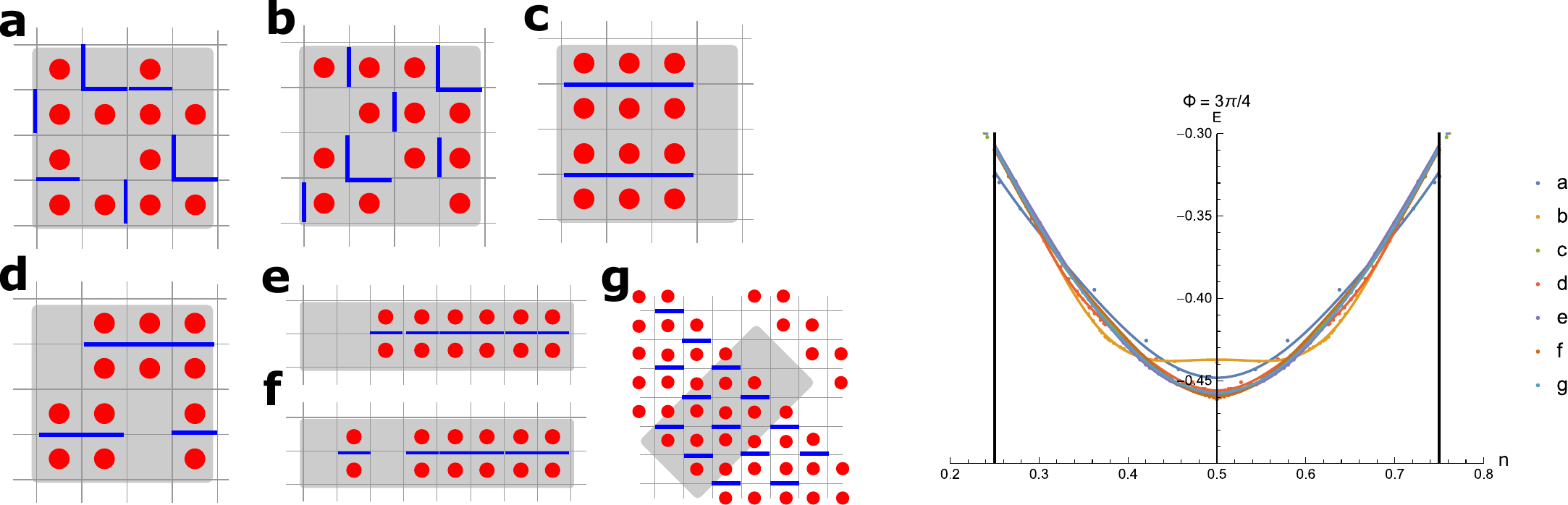}
\caption{Configurations of average flux $\Phi = 3\pi/8$ and associated fermionic ground state energies as a function of filling.}
\label{fig:Configs3Pi8}
\end{minipage}

\begin{minipage}{\textwidth}
\includegraphics[scale=.4]{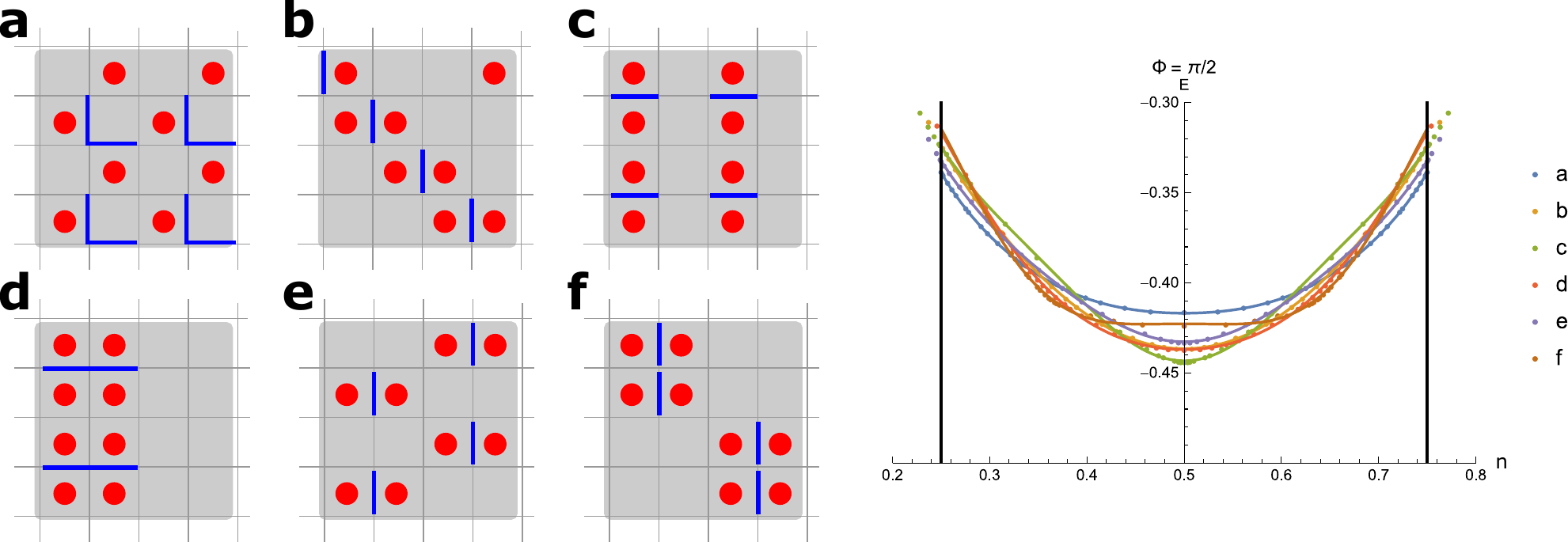}
\caption{Configurations of average flux $\Phi = \pi/2$ and associated fermionic ground state energies as a function of filling.}
\label{fig:ConfigsPi2}
\end{minipage}
\end{figure}

\begin{figure}
\begin{minipage}{\textwidth}
\includegraphics[scale=.4]{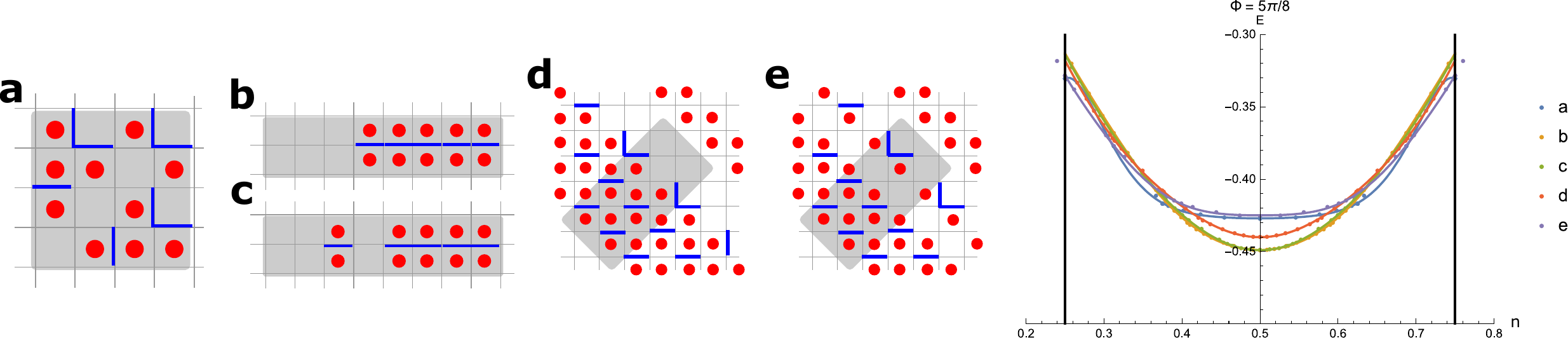}
\caption{Configurations of average flux $\Phi = 5\pi/8$ and associated fermionic ground state energies as a function of filling.}
\label{fig:Configs5Pi8}
\end{minipage}

\begin{minipage}{\textwidth}
\includegraphics[scale=.4]{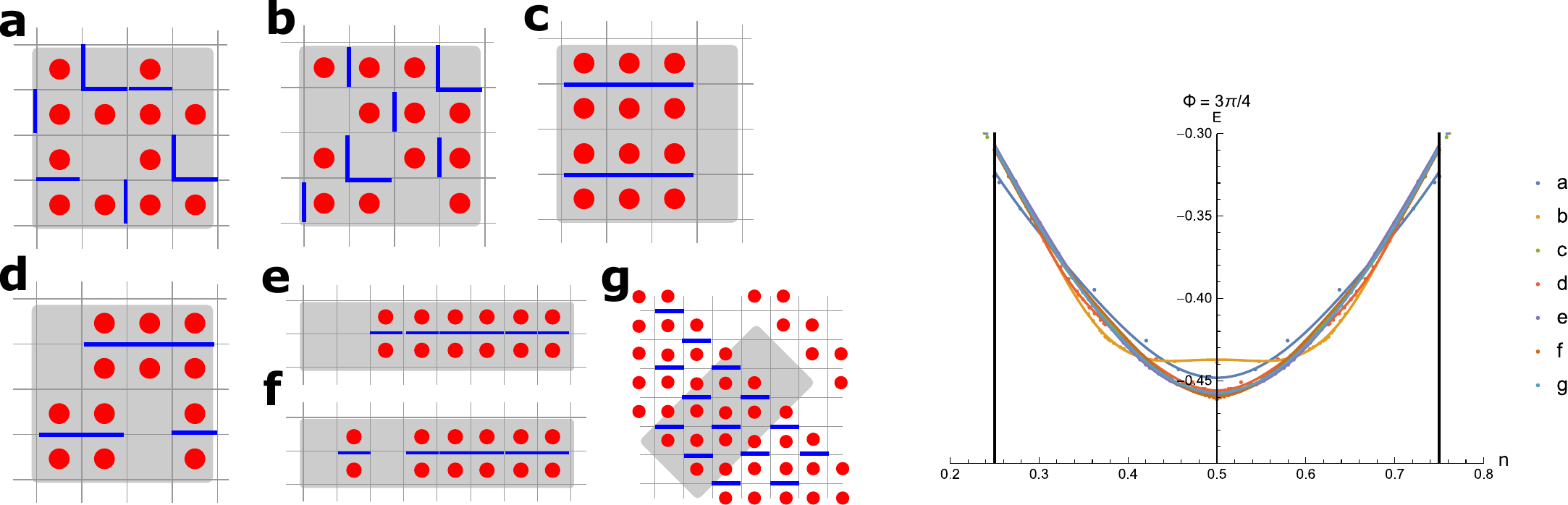}
\caption{Configurations of average flux $\Phi = 3\pi/4$ and associated fermionic ground state energies as a function of filling.}
\label{fig:Configs3Pi4}
\end{minipage}

\begin{minipage}{\textwidth}
\includegraphics[scale=.4]{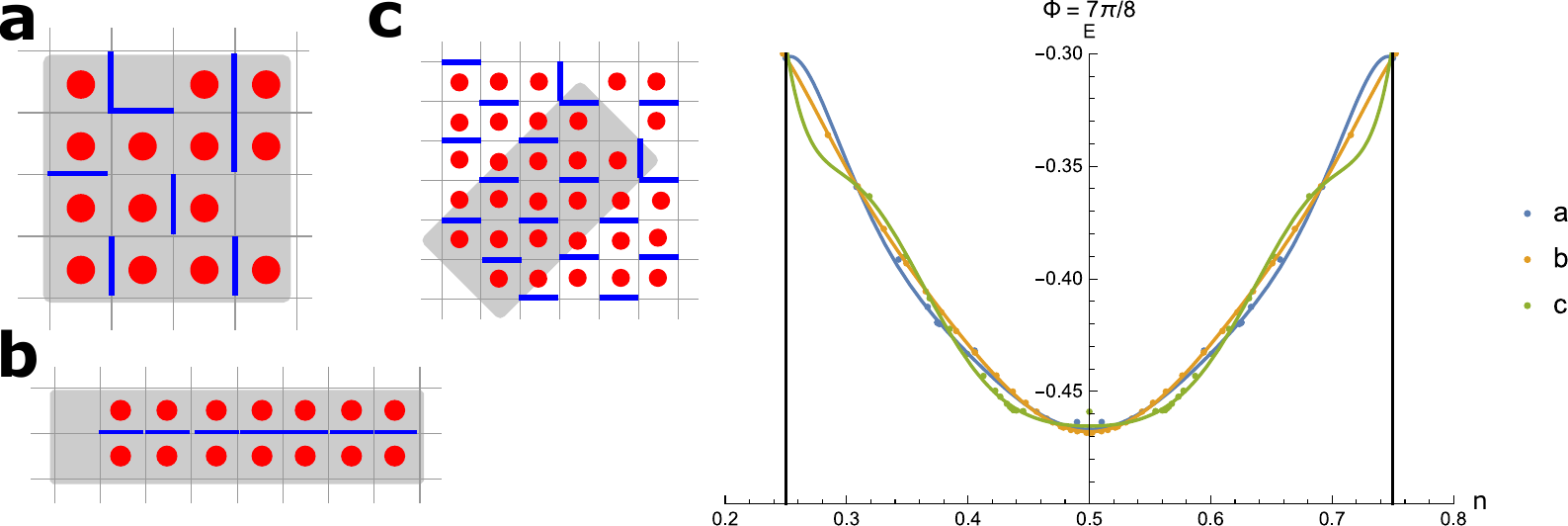}
\caption{Configurations of average flux $\Phi = 7\pi/8$ and associated fermionic ground state energies as a function of filling.}
\label{fig:Configs7Pi8}
\end{minipage}
\end{figure}
\end{widetext}
\newpage

\bibliography{Z2gauge}

\end{document}